\documentclass[aip,pof]{revtex4-1}
\usepackage{graphicx}
\usepackage{amsmath}
\usepackage{amsfonts}
\usepackage{amssymb}
\usepackage{epsf}
\usepackage[usenames,dvipsnames]{color}

\usepackage{comment}
\newcommand{\beq}{\begin{equation}}
\newcommand{\eeq}{\end{equation}}
\def\beqa#1\eeqa{\begin{align}#1\end{align}}
\def\bcorr #1 \ecorr{ {\color{red} #1 } }
\newcommand{\bu}{ \mathbf u }
\newcommand{\bp}{ \mathbf p }
\newcommand{\br}{ \mathbf r }
\newcommand{\er}{ \mathbf e_r }
\DeclareMathOperator{\bF}{\bf F}
\newcommand{\grad}{\boldsymbol{\nabla}}

\newcommand{\divg}{\text{div}}
\newcommand{\mcl}{\mathcal}
\DeclareMathOperator*{\dt}{\partial_t }
\DeclareMathOperator*{\dr}{\partial_r }
\DeclareMathOperator{\bP}{\bf p}
\DeclareMathOperator{\bI}{\bf I}

\DeclareMathOperator{\delu}{\delta{\mathbf u}}
\DeclareMathOperator{\deln}{\delta n}
\DeclareMathOperator{\delP}{\delta p}
\DeclareMathOperator{\delR}{\delta R}
\DeclareMathOperator{\delur}{\delta u_r}
\DeclareMathOperator{\deluT}{\delta \mathbf{u}_{\perp}}
\DeclareMathOperator{\delVT}{\delta V_{\!\perp}}
\DeclareMathOperator{\delVr}{\delta V_r}
\DeclareMathOperator{\delM}{\delta M}
\DeclareMathOperator{\delwP}{\delta  P}
\DeclareMathOperator{\be}{\bf e}
\DeclareMathOperator{\gradT}{ \boldsymbol \nabla_{\!\! \perp}}

\newcommand{\DR}[1]{{\color{ForestGreen}#1}}
\begin{document}
\title{Microscopic origin of self-similarity in granular blast waves}

\author{M. Barbier}
\affiliation{Department of Ecology and Evolutionary Biology, Princeton University, Princeton, NJ 08544, USA}

\author{D. Villamaina}
\affiliation{Laboratoire de Physique Th\'eorique de l'ENS (CNRS UMR 8549) and Institut de Physique
Th\'eorique Philippe Meyer, 24 rue Lhomond 75005 Paris - France}

\author{E. Trizac}
\affiliation{LPTMS, CNRS, Univ. Paris-Sud, Universit\'e Paris-Saclay, 91405 Orsay, France}

\date{\today}

\begin{abstract}
The self-similar expansion of a blast wave, well-studied in air, has peculiar counterparts in dense and dissipative media such as granular gases. 
Recent results have shown that, while the traditional Taylor-von Neumann-Sedov (TvNS) derivation is not applicable to such granular blasts, 
they can nevertheless be well understood via a combination of microscopic and hydrodynamic insights. In this article, we provide a detailed 
analysis of these methods associating Molecular Dynamics simulations and continuum equations, which successfully predict hydrodynamic profiles, 
scaling properties and the instability of the self-similar solution. We also present new results for the energy conserving case, including the particle-level analysis 
of the  classic TvNS solution and its breakdown at higher densities.

\begin{comment}
 Dissipative collision cascade ;
dynamics of a localized intense explosion ; strong shock ; particle /
MD simulations ; hydrodynamics supplemented with scaling analysis ;
self-similarity ; structure of blast ; instability.
\end{comment}
\end{abstract}


\maketitle

\section{Introduction}
A blast is a shock wave that follows the sudden release of a large amount of energy in a small volume of gas -- typically due the detonation of an explosive. 
The volume of gas is heated to large temperatures, thus large pressures, and starts expanding, causing a decrease of density at the center and a corresponding increase toward the 
boundaries. In the case of ``strong shocks", the displaced matter moves faster than energy can be transported into the environment by sound or heat waves, and there is thus a discontinuity between a well-defined expanding perturbed zone and the surrounding gas, still at rest. This compression shock front is characterized by a sharp gradient in density and other hydrodynamic fields. 

In air and similar conservative gases, it is well-known that as long as this strong shock condition is verified, the expansion of the bulk of the blast is self-similar in time: it exhibits a fixed internal structure, which simply scales up over the course of the dynamics. Furthermore, all the scaling laws for observables are easily derived from dimensional analysis. This in turns allows for greatly simplified, even solvable theoretical descriptions  \cite{Taylor1950,Taylor1950b}, and has been studied experimentally in a number of systems, from air~\cite{dewey1964air} to laser-induced shocks in plasmas~\cite{Edens2004,Moore2005} or astrophysical systems such as supernova remnants~\cite{Cioffi1988} (the interaction of ejecta from stellar explosions 
with the surrounding environment). 

\begin{figure}
\centering
\includegraphics*[width=350pt,clip=true]{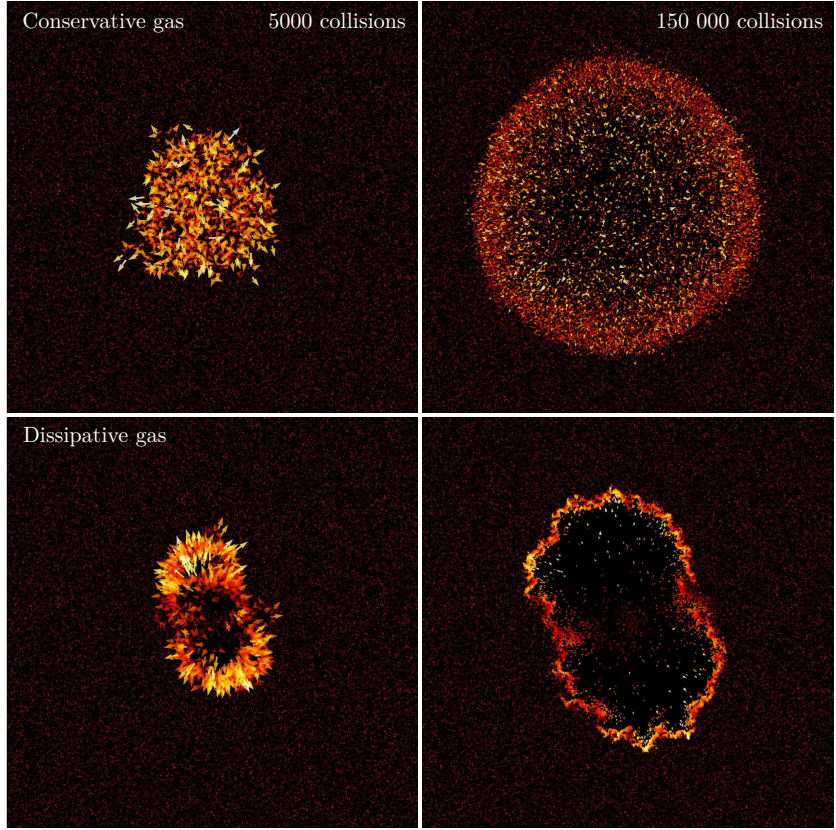}
\caption{Snapshots of a self-similar blast in Molecular Dynamics simulations, with particles indicated by disks and their velocities by arrows. 
The figure shows the collision cascade after $5\,10^3$ and $1.5\,10^5$ collisions, starting with two particles at the center with equal and opposite initial velocities. 
The conservative (top row) and dissipative (bottom row) cases differ in a number of ways; most strikingly, the conservative blast restores its rotational symmetry while anisotropy is preserved by dissipative dynamics, which also lead to a hollow core devoid of particles. \label{fig:cover} }
\end{figure}

Many  applications however diverge from the prototypical example of air, by involving media that dissipate energy or momentum due to radiation, inelastic collisions or drag~\cite{Ostriker1988,Barenblatt1996}, and may also exhibit high or inhomogeneous density and other unusual features. In many cases, self-similar expansion can still be observed, although scaling properties can generally no longer be explained by mere dimensional analysis. A striking counter-example to the latter is the asymptotic scaling regime of the blast in an energy-dissipating medium such as a radiative plasma or granular gas. As seen in Fig.~\ref{fig:cover}, a blast wave in this setting takes a peculiar hollow shell structure, whose expansion can be entirely understood in terms of inertial motion.

Across this wide range of problems, approaches can broadly be divided into three categories, depending on the level of description that they adopt:
\begin{itemize}
\item Scaling laws for macroscopic quantities such as the radius $R(t)$ and the energy $E(t)$, derived from considerations on conservation laws and fluxes at the boundaries, e.g.~\cite{Ostriker1988}.
\item Continuum media descriptions: hydrodynamic models and finite-element (or analogous) simulations, e.g.~\cite{Cavet2008,Sari2011}.
\item Particle-level descriptions: kinetic theories and Molecular Dynamics simulations,~e.g. \cite{Antal2008}.
\end{itemize}
A thorough understanding of self-similarity may often require the joint use of all three types of approaches. For the dissipative blast in particular, it has long been suggested that the global invariant driving this similarity regime is conserved radial momentum~\cite{Oort1951}; however, a microscopic justification for how this regime comes about and is maintained has been lacking, and the previous attempt at a continuum description has not been empirically validated~\cite{Barenblatt1996}.
 Unfortunately, the latter two types of approaches, coarse-grained and microscopic, have so far remained largely impervious to each other. 

Following the success of Taylor~\cite{Taylor1950b} and others, blasts have long been associated with continuum descriptions, which comprise the vast majority of the literature on this topic~\cite{Ostriker1973,Ostriker1988,Cioffi1988}. However, some of the assumptions underlying these analyses, and even the associated simulations, are questionable in the absence of microscopic scrutiny. Indeed, the derivation of hydrodynamic equations, and constitutive relations for pressure and transport terms, assumes local thermodynamic equilibrium conditions at each point of the medium. This can fail in shockwaves~\cite{colangeli2013kinetic}, especially within the shock front, traditionally considered as a singular boundary in continuum equations, where hydrodynamic fields undergo large changes over very small distances. These assumptions are also problematic in non-conservative fluids. Rigorous derivations from a microscopic perspective are mostly missing, and rarely if ever connected with the existing body of knowledge on blasts.

On the other hand, particle-level approaches have only recently turned to the question of blast waves~\cite{Antal2008}.
This initiative mostly stems from the study of granular gases, which has opened a new arena in shock wave dynamics~\cite{Brill}: these fluids of macroscopic grains provide a direct window into kinetic phenomena, and have been offered as a prototype for understanding energy-dissipating media~\DR{\cite{puglisi2014transport}}. As we will show, this is especially true in the case of the dissipative blast:
under weak conditions, its self-similar growth does not depend on the specifics of the dissipation mechanism, and one can confidently use a granular gas as a model system for other dissipative fluids, allowing for a detailed grasp of phenomena that may also occur in laboratory plasmas and astrophysical systems.
Blast experiments in granular fluids~\cite{Boudet2009,boudet2013unstable} and particle based simulations \cite{Jabeen2010} have been performed in recent years, giving snapshots such as those in Fig.~\ref{fig:cover} into the peculiar structure of the shock. They have now been successfully analyzed by a continuum description~\cite{letter}. This multilevel approach  is explained in detail here, and applied back to the original problem of conservative fluids, leading to the first Molecular Dynamics validation of the Taylor model, its extension to a more general constitutive relation, and the discovery of its breakdown at higher densities. 
This approach calls for replication for other self-similar phenomena driven by global invariants, that may likewise find relevance beyond granular systems.

The plan of this article is as follows: in Sec.~\ref{sec:state}, we recall previous results on blast 
waves, their self-similarity properties, and the study of shock waves in the granular fluid literature.
Then, we expound on the hydrodynamic framework in Sec.~\ref{sec:hydro}: first for a conservative blast according to the Taylor-von Neumann-Sedov approach, which we generalize to higher densities and confront to simulation results; then in the presence of dissipation, which requires a new approach based on a multi-layered structure. From there, we explain the peculiar scaling of the dissipative blast in Sec.~\ref{sec:scal}. Finally, we describe in Sec.~\ref{sec:instab} the mechanisms 
through which the previously derived hydrodynamic solution becomes asymptotically unstable, and we explain the self-similar growth of this corrugation instability.
Preliminary accounts of parts of this work were published in \cite{letter}.

\section{State of the art}
\label{sec:state}

\subsection{Taylor-von Neumann-Sedov (TvNS) theory}

\label{sub:TvNS}

Self-similar shock waves in molecular gases were extensively studied over the 1940-1950 decades, toward the practical concern of describing the blast wave caused in air by the detonation of a nuclear weapon. The usual pictures associated with such bombs are those of the initial flash and the fireball, made of ignited air and debris which expands while subjected to strong convection, and gives rises to the prototypical mushroom shape. However, a large fraction of the damage caused by the explosion actually comes from the blast: the surrounding medium is brought to very high temperature and pressure, and thus set into fast expanding motion~\cite{glasstone1964effects}. At this stage, nuclear weapons differ from conventional explosives only by the magnitude of the overpressure and the velocity of the gas flow. 
Nonetheless, this difference has deep consequences for the behavior of the blast: contrary to lesser explosions, the outward velocity of the wind (matter displacement) is larger than that of sound or 
heat waves in the external medium. Were this not the case, either type of waves could transport some of the energy of the explosion outward, and progressively attenuate the difference of pressure between 
the fluid inside and outside the blast. By contrast, supersonic adiabatic blasts are characterized by a sharp transition into an expanding high-pressure region, and this boundary or ``shock front'' 
remains singular until its decreasing velocity ceases to meet the above conditions. This compression at the front and winds within the blast are responsible for much of the damage caused by the weapon.

This property is crucial for the description of the blast, as it means that two conservation laws apply within the same radius $R(t)$: on one hand, the total number $N(t)$ of particles within the blast, or their cumulative mass $M(t)$, must equal the one found in the same region before the perturbation; on the other hand, the total energy initially imparted by the explosion remains contained within that region $E(t)=E_0$. 
As alluded to in the introduction, Sir~G.I.~Taylor famously used these arguments to deduce the scaling law for $R(t)$, using one further assumption that we now explain. 
The energy $E(t)$ has two contributions: one from motion in the radial direction, with an average velocity that must be proportional to $\dot R(t)$, and one from undirected motion 
(thermal agitation) within the blast region. Transport effects such as viscosity convert some of the energy of coherent flow into agitation, but assuming that the fraction 
of energy involved in expansion is non-vanishing, the following scaling holds asymptotically
\beq  \dfrac{ M(t) \dot R^2(t)}{E(t)} \sim 1. \eeq
Since we further know that the mass of the blast region is
\beq M(t)\sim \rho_0 R^d(t) \eeq
with $\rho_0$ the mass density of the medium and $d$ the dimension of space, and that energy is conserved, it comes naturally that
\beqa E(t) &\sim R^d(t)\; \dot R^2(t) \sim E_0, \eeqa
hence
\beqa
\dfrac{d}{dt} R(t) &\sim R^{-d/2}(t),\\
R(t) &\sim t^{\frac{2}{d+2}}.
\eeqa
The same law can be found from dimensional analysis: the problem supposes four quantities $R$, $t$, $E_0$ and $\rho_0$ which exhibit only three independent dimensions, therefore we can construct a single dimensionless ratio, 
\beq  g=  \dfrac{E_0 t^2}{\rho_0  R^{d+2}(t) }, \label{eq:defg} \eeq
or some power thereof.
Taylor gave physical arguments to justify that $g\approx 1$ in real systems~\cite{Taylor1950} and was thus able to compute $E_0$ with good accuracy for the Trinity test, 
knowing only $\rho_0$ for air and a few values $R(t)$ from some publicly available snapshots of the blast. More generally, after rescaling, any blast is only characterized by $g$ without any explicit time dependence; thus, this ratio is a constant of motion for a given blast, again leading to  $R(t) \sim t^{\frac{2}{d+2}}$. This agreement between dimensional analysis and explicit calculations involving conservation laws is far from coincidental, as we explain in the next section.  
It is also appropriate to mention that the blast mass $M(t)$ shows a power-law growth in $R(t)^d \sim t^{2d/{(d+2)}}$,
which is the mean-field prediction for the ballistic coalescence model \cite{TrHa95,TrKr03}. 
It can be shown that an exponent $2d/(d+2)$, first predicted in \cite{CaPY90}, while a rather 
poor approximation for the original model, becomes exact in the present context. Here indeed, the system behaves as a sticky gas:  
a {\em single} agglomerate grows in an environment of particles at rest and when a particle collides, it merges with it, validating the 
scaling arguments presented in~\cite{CaPY90}. 

%\ET{XXXX Btw, it would be funny to see what would happen with interacting blasts : generate several blasts
%with several seeds, and see how their merging operates ; already, how do two such blasts interact ?
%May be quite demanding computationnaly.}

Going beyond simple scaling laws, the spatial structure of the blast was independently derived by Taylor in England~\cite{Taylor1950b} and two other luminaries: von Neumann in the United States \cite{Neumann1944} and Sedov in the Soviet Union \cite{Sedov1946}. Rather than the global quantities $E(t)$, $M(t)$ and $\dot R(t)$ that characterize average properties over the whole perturbed region, they considered the state of the gas within concentric regions of radius $r<R(t)$, thus switching to a continuum description of the blast.
The dimensional argument above still holds: since physical laws can only involve dimensionless parameters, and assuming that the blast is isotropic (exhibiting a central symmetry), the value of any hydrodynamic field at a given point of space depends only on a single variable
\beq \lambda = \dfrac{r}{R(t)} = r  \left( \dfrac{g \rho_0}{E_0 t^2 } \right)^{1/(d+2)} \eeq
where we used Eq.~\eqref{eq:defg} to rewrite $R(t)$ in terms of other dimensional quantities and the characteristic constant $g$.
The distribution of matter and energy within the blast is therefore self-similar with respect to time, and blasts with different values of $g$ can be made to correspond to 
concentric slices of the same structure. 
Inserting this scaling ansatz in hydrodynamic equations gives an elegant, exactly solvable model that has been exhaustively validated in dilute, conservative fluids such as air \cite{Landau1987fluid}.  Recalling this textbook result is beyond the scope of the present article, and only the aspects relevant to our problem will be expounded upon in Section~\ref{sec:hydro}. But it is worth emphasizing that all scaling relationships in the  Taylor-von Neumann Sedov (hereafter TvNS) theory are independent of any microscopic detail: they can be derived immediately using dimensional analysis or, equivalently, the existence of global invariants. This qualifies the TvNS blast as an example of self-similarity \textit{of the first kind} (see below). It has thus become the model for all studies of self-similar shockwaves, and a classic illustration of the general theory of self-similarity and dimensional analysis \cite{Barenblatt1996}.

\subsection{Self-similar blasts of the first and second kind}
\label{sub:selfsimkind}
Numerous variations on the TvNS blast have been considered over the past decades, mostly by relaxing some of the constraints (conservation laws) that shape the prototypical solution. We will focus here on other strong shocks: weak shocks, expanding with velocities comparable or inferior to the sound or heat wave velocity in the external medium, tend to have no well-defined boundary; therefore, they do not usually exhibit the conspiration of global conservation laws over the same circumscribed domain which ensures the self-similarity and simplicity of the TvNS solution.

A common extension is to relax the energy conservation  within a  limited region, usually by considering energy production or absorption either at the center or at the boundary of the shock. The former is relevant to problems with a permanent source rather than an initial discharge of energy. The latter can represent exo- or endothermic chemical phenomena such as flame or other reaction fronts, where it is assumed that the media on both sides of the boundary are non-reactive. In most cases, all conservation laws are still satisfied in the bulk of the flow; this usually ensures that the growth of the  perturbed domain remains self-similar, but the scaling exponents for global quantities cannot be determined by dimensional analysis anymore: the expansion speed is generally controlled by the phenomena happening within the front, and some dynamical description, or techniques such as renormalisation, are required to determine the exponents which depend continuously on dynamical parameters.
 Barenblatt, in his extensive and pedagogical exposition of scaling phenomena \cite{Barenblatt1996}, thus characterized two classes of self-similarity: the \textit{first kind}, exemplified by the TvNS solution, where only global conserved quantities are relevant variables (to borrow a term from renormalization theory) and the appropriate dimensional analysis is sufficient to extract all scaling properties; and the \textit{second kind}, in which some dimensionless combinations of microscopic parameters remain relevant to the dynamics and enter the exponents. Many critical phenomena, such as directed percolation, can be seen to belong to this second class.

An important example of both kinds of self-similarity is found in  blast waves with energy dissipation in the bulk rather than on the shock boundary, such as our granular blast. Indeed, conjectures on the type of scaling laws found in such systems date as far back as Oort \cite{Oort1951}. He proposed that their 
asymptotic regime is one where all the matter in the shocked region is condensed into a thin hollow shell which propagates only under the force of its conserved momentum, and decelerates due to continual accretion of matter from the outside. This regime, traditionally described in terms of a singular boundary layer containing all the accreted matter, is known as the Momentum-Conserving Snowplow (MCS), and it is self-similar of the first kind: the typical momentum in the radial direction $\Pi(t)$ is conserved and scales as
\beq \Pi(t)  \sim N(t) \dot R(t) \sim t^0 \hspace*{50pt}  \Leftrightarrow\hspace*{50pt}  R(t)\sim t^{\frac{1}{d+1}} \label{eq:MCS-history} \eeq
An intermediate regime has been proposed later under the name of Pressure-Driven Snowplow \cite{McKee1977}: it is thought to occur before the MCS, when most matter has already formed a shell but  dilute hot gas remains at the center and pushes the shell outward. Assuming that dissipation in that inner pocket of gas is negligible due to its low density, and thus that it expands adiabatically, it is easy to show (more details in section \ref{sub:regimes}) that
\beq P(t) R(t)^{d\gamma} \sim t^0  \hspace*{50pt}  \Leftrightarrow\hspace*{50pt}  R(t)\sim t^{2/(d+\gamma+2)} \label{eq:PDS-history} \eeq
which is thus an example of self-similarity of the second kind: the microscopic dimensionless parameter $\gamma$ (the adiabatic index of the gas) appears in the exponent. 
This regime is expected to be the only asymptotic one in cases where the dissipation rate decreases with increasing temperature (contrary to granular flows), and the center thus always remains hotter than the periphery \cite{Cioffi1988}.

These two regimes have been proposed in the case of a related problem in astrophysics and plasma physics  \cite{Ostriker1988}, the so-called \textit{radiative} blast: such shockwaves occur mostly through supernovae causing a displacement of the interstellar medium, where kinetic energy is mostly lost by being converted to radiation in frequency ranges where the medium is transparent, and thus cannot reabsorb it. Commonly studied mechanisms include inverse Compton scattering, Bremsstrahlung, and dust cooling \cite{Ostriker1973}; the latter is in fact kinetic in nature as it is caused by collisions with suspended mesoscopic grains, and thus has strong similarities with the dissipation encountered in granular flows.

We should mention one last branch of related studies, which investigate the limits of the original TvNS model for a conservative blast in more exotic media. Some alter the relationship $M(R)$ by having density before the blast be a power-law of the form $\rho_0(r)\sim r^\Omega$ \cite{Book1994,Chevalier1982}, as can occur in astrophysical systems, where the repartition of matter around the supernova-to-be may result from some previous phenomenon with central symmetry. This has the consequence that $R(t)\sim t^{\frac{2}{d+2-\Omega}}$, allowing to "simulate" different values of the spatial dimension $d$, with an obvious change of regime as $\Omega \to d+2$. 
This can play a significant role in the case $d=1$, which is generally considered pathological for the conservative blast~\cite{blast1d}.
Another series of works focuses on gases with peculiar values of adiabatic index $\gamma$, which does not usually intervene in the scaling laws, but rather in the hydrodynamic profiles. A significant result is the observation that the TvNS solution destabilizes as $\gamma \to 1$ (corresponding to molecules with a high number of internal degrees of freedom). This prompted further investigation of instabilities in self-similar blasts \cite{Ryu1987,Ryu1991}. The latter direction is especially interesting for us as the granular blast exhibits an instability that is not reducible to any of those previously studied, as explained in section \ref{sec:instab}.

\subsection{Granular shocks}
\label{sub:granushocks}

Finally, the granular blast itself has recently started to attract studies, both experimental \cite{Boudet2009,boudet2013unstable,Vilquin} and numerical \cite{Jabeen2010,Pathak2012,Pathak2012b}. These papers have been focusing on global scaling laws, revealing clear self-similarity in a MCS regime (see previous section) in Molecular Dynamics simulations

Another concern in these papers is a particle-level description of the blast,
as the macroscopic size of the grains allows for precise experimental measurement of kinetic properties. Of particular interest is the shock front, which can only be represented as a singular boundary layer in hydrodynamic frameworks: granular systems offer a unique opportunity to analyse their structure and check the assumptions usually made in the study of molecular systems. Therefore, the aforementioned experiments  \cite{Boudet2009,boudet2013unstable} belong to a corpus of mostly empirical and numerical studies on shocks in other types of granular flow: nonlinear acoustic waves and solitons \cite{Bougie2002, Rosas2003} with analytical unidimensional models \cite{Nesterenko1984,Daraio2005,Sen2008}, and shocks caused by an obstacle in the flow \cite{Horluck1999,Rericha2001,Gray2007,Boudet2008}.

Although we do not build a kinetic theory of the shock front, and only account for it through the boundary conditions it imposes on the flow, these works are relevant to our problem. They outline the necessity of proving that our simplified description is quantitatively sufficient and does not fail due to neglected microscopic phenomena.

\section{Hydrodynamics}
\label{sec:hydro}

\subsection{Framework}
Previous approaches of the dissipative blast have been rooted in the study of astrophysical or laboratory plasmas, and thus rely on hydrodynamic descriptions for these fluids.

The study of granular gases lends itself naturally to a kinetic approach and Molecular Dynamics simulations. It is however possible to connect those to a hydrodynamic framework, by coarse-graining the particle-level description so as to define their density $n(\br,t)$ (locally averaged over realizations or with an appropriate smoothing kernel), as well as the mean $\bu(\br,t)$ and variance $\Theta(\br,t)$ of their local velocity distribution. Equations thus obtained from kinetic theories and coarse-graining have the structure of hydrodynamic models, with coefficients expressed in terms of particle properties; they were first derived in the limit of dilute systems \cite{brey1998hydrodynamics,JeSa83}, then extended to dense fluids of inelastic hard spheres \cite{GaDu99}. 

An important simplification in our case is that, since we are interested in a self-similar solution with growing length scale $R(t)$, spatial gradients tend to decrease in magnitude as for any field $\psi$
\beq \dfrac{d}{dr}\, \psi\!\left( \dfrac{r}{R(t)} \right) = \dfrac{1}{R(t)} \dfrac{d}{d\lambda} \psi(\lambda). \eeq
Thus, terms of different order in spatial gradients will scale differently with $R(t)$ and decouple asymptotically. This property is especially interesting as the derivation of hydrodynamic equations from a kinetic framework involves an expansion in powers of the gradients, the Chapman-Enskog expansion~\cite{Chapman1991mathematical}, which is only formal in most systems yet becomes exact in the bulk of the blast. More concretely, transport terms such as viscosity and heat conduction, involving higher-order gradients, cannot manifest at the same spatial scale as advection, i.e. $R(t)$; thus, we will see that the flow at that scale essentially obeys the Euler equations for a perfect fluid. The consequences of considering heat conduction have been discussed in the literature and shown to hold only for a limited time~\cite{Reinicke1991}.

Up to the first order in gradients, the equations for the local evolution of mass, momentum and energy in \cite{GaDu99} take the standard form of compressibility, Navier-Stokes and heat equations~\cite{Landau1987fluid}
\begin{subequations}\label{eq:hydro}
\beqa & \dt n + \nabla (n\bu) = 0  \\
& (\dt + \bu.\nabla) \bu + \dfrac 1 n \nabla.\bP =0  \label{eq:uhydro}\\
&  n(\dt + \bu.\nabla) \Theta +  (\gamma-1) (\bP.\nabla).\bu  =-\Lambda  \label{eq:Ehydro}
\eeqa
\end{subequations}
where the adiabatic index is given by $\gamma =1+\frac{2}{d}$ for hard spheres in spatial dimension $d$.
The only term specific to dissipative fluids is the  energy sink $\Lambda$ given by 
\beq \Lambda = \omega (1-\alpha^2) n\Theta
\label{eq:dissip} %-3/2 (1-\alpha) \,(1-Z^{-1}(n)) (\bP.\nabla).\bu
\eeq
where $\omega = \omega_0 (n\Theta)^{1/2}$ is the collision frequency in the gas, proportional to the average relative velocity and the inverse of the mean free path \cite{Goldshtein1996,GaDu99}, with $\omega_0$ a dimensionless constant. The constant $\alpha$ is known as the restitution coefficient, and $(1-\alpha^2)$ represents 
the average fraction of thermal energy lost in each collision: if $\alpha=1$, collisions are elastic, and if $\alpha=0$, they are maximally dissipative. 
It is known in granular systems that $\alpha$ for any collision may in fact be a function of the relative velocity of particles \cite{Poschel2003}, but here we will show that any $\alpha<1$ will lead to the same asymptotics, and we can thus assume $\alpha$ to be constant without loss of generality.

A similar energy sink term occurs in the description of radiative plasmas, where it is expected to exhibit some power-law dependence in $n$ and $\Theta$
\beq \Lambda = \Lambda_0 n^{\mu+1} \Theta^{\nu+1} \eeq
with $\Lambda_0$ an appropriate dimensional constant. The energy dissipation rate, having the dimensions of an inverse time, is thus given by $\Lambda_0 n^{\mu} \Theta^{\nu} $.  When the origin of dissipation is collisional, as in granular flows or dust cooling in interstellar media (the plasma losing energy through collisions between ions and dust particles),  we expect $\mu=\nu=\frac 1 2$ \cite{Ostriker1973}. 
Indeed, the r.h.s. of Eq.~\eqref{eq:dissip} can be rewritten as $\Lambda_0(\alpha)\, (n\Theta)^{3/2}$ where we identify $\Lambda_0(\alpha)=\omega_0 (1-\alpha^2)$. 
A very different behavior can be expected to arise depending on the sign of $\mu$ and $\nu$ \cite{Bertschinger1986}. 
However, if both are positive -- as is the case here -- meaning that the dissipation rate increases with density and with agitation, we argue that the solution is largely 
independent on the specific values of these exponents: 
the same blast structure (described below) will  be established for any $\mu,\nu>0$, as the energy sink then exceeds the advection terms in Eq.~\eqref{eq:Ehydro}, 
and all temperature is dissipated within a comparatively thin cooling layer. The exponents $\mu$ and $\nu$ may control the precise features of that layer, 
but they will not alter the scaling and stability properties of the blast, which we show below are all driven by the inner cold region where motion is strictly coherent.

\subsection{Constitutive relation}

The hydrodynamic equations \eqref{eq:hydro} are closed by specifying the pressure tensor ${\bP}$ with a constitutive relation. It may be assumed to have vanishing non-diagonal terms:  those would contribute to transport phenomena (i.e. viscosity) involving higher-order gradients.  As argued above, those are irrelevant here because their scaling in the growing length scale $R(t)$ will generically differ from that of the terms in Eqs.~\eqref{eq:hydro},  causing asymptotic decoupling. 

In hydrodynamic models, the diagonal component of pressure is commonly assumed to be isotropic, i.e. $\bP=p \bI$ with ${\bI}$ the identity tensor. This is a natural consequence of the existence of a local equilibrium at every point in the fluid, which underlies standard hydrodynamics. However, this assumption does not necessarily hold in granular systems, or dissipative fluids in general, and indeed will be challenged below. Instead, we propose that a solution can be found with
\beq \bP=p\, \textbf{e}_p \eeq
where $\textbf{e}_p$ is a tensor reflecting the directionality of pressure and will be specified as needed. As for the scalar magnitude $p$, the dense transport framework for hard spheres~\cite{GaDu99} suggests the constitutive relation
\beq p = n \Theta Z(n) \eeq
where $Z(n)$ stems from the finite compressibility of the gas. In the dilute limit, $Z(n \to 0)\to 1$ and the relation above becomes the  ideal gas law (note that $k_B$ does not appear as $\Theta$ is an energy rather 
than a thermodynamic temperature). However, $Z(n)$ diverges at finite $n$ to account for the increase of pressure due to steric effects. Numerous functional forms have been proposed 
for $Z(n)$~\cite{Baus1987}; for our two-dimensional study, we choose the classic Henderson relation \cite{Henderson1975}
\beq Z(n) = \dfrac{1+\varphi(n)^2/8 }{(1-\varphi(n))^2} \label{eq:HendersonZ} \eeq
with the volume fraction $\varphi(n)$ as defined in \eqref{eq:defphi}. This relation is derived under equilibrium conditions; yet, it is found to provide good agreement between theoretical and numerical results, even as it is used here to describe the strongly out-of-equilibrium flow in the blast wave, and especially within the shock front. 
Other plausible candidates can be found in \cite{Santos1995} and were found to provide comparably good results.

This specification of the constitutive relation is an important step in our analysis, as we show below that even in the well-known conservative blast, numerical results at moderate densities can disagree significantly with the standard TvNS solution. This discrepancy is a new result, and will also provide an occasion to recall steps of the standard method that prove useful for our discussion of the dissipative blast.

We must additionally remark that with a nontrivial constitutive relation $Z(n)\neq 1$, another term enters in the expression of the energy dissipation, which is now properly written as
\beq \Lambda =\Lambda_0(\alpha)\, (n\Theta)^{3/2}  \;-\; \Lambda_1(\alpha) \,(1-Z^{-1}(n))(\bP.\grad).\bu\eeq
The second term is seen to be structurally similar to one on the left-hand side of Eq.~\eqref{eq:Ehydro}: while the latter signifies random motion created by pressure along a velocity gradient, this term represents the energy dissipated in the same conditions; the two  can be written as one by using an effective adiabatic index $\gamma^*(n,\alpha)$.
Various expressions have been proposed for the prefactor $\Lambda_1(\alpha)$~\cite{Goldshtein1996b,Pagonabarraga2001}. While this new term should be included to achieve a consistent level of approximation throughout the equations, we found that it did not alter the results perceptibly, and ignored it for simplicity.

\subsection{Shock front and Rankine-Hugoniot conditions}
\label{sec:RH}
Once the hydrodynamic equations are closed by the choice of an appropriate equation of state, a solution for the fields is entirely specified by its boundary conditions. In our case, the natural boundary is the shock front: the layer separating the bulk of the blastwave from the external medium. It has a clear microscopic definition, as the region within which excited particles mix and collide with particles at rest, creating incoherent motion ("temperature"). The radius of the blast $R(t)$ can thus be defined either as the inner limit of the shock front, so as to cover only excited particles, or as its outer limit, so as to exclude only particles at rest; we will opt for the former definition so that all particles within the blast are excited, which for instance allows this whole region to be self-similar in the TvNS solution.

While the shock front may in principle be described in a kinetic theoretical approach \cite{Boudet2008}, it is singular from the point of view of hydrodynamics: its width  $\epsilon$ is microscopic, of order a few mean free paths in the gas at rest, and the velocity distribution of particles in the front is multimodal as some are excited and other immobile (and yet other fall in-between). By contrast, hydrodynamic equations typically assume a local equilibrium, i.e. a velocity distribution close enough to the gaussian to be correctly described by its mean and variance only.

The usual approach  at the continuum level is to treat the front as a singular surface, and compute only its in-bound and out-bound fluxes, to derive  boundary conditions on the fields in the blast. While the  hydrodynamic equations Eqs.~\eqref{eq:hydro} do not necessarily hold at each point in the front, they are only one local embodiment of more general conservation laws. A useful perspective is then to formulate these conservation principles for fluxes of matter, momentum and energy going through the front: ignoring all details 
of what happens within, no matter or momentum can disappear, nor energy in the case of elastic collisions.

We assume once again that orders in spatial gradients are decoupled (this time because the length scale $\epsilon$ is considered as infinitesimal and thus terms with higher-order gradients would have much larger magnitude). The out-of-equilibrium nature of the front can affect only two expressions: the constitutive relation for $\bP$, and the energy sink term $\Lambda$. We will assume here that the bulk expression can be retained for both of them, then discuss the discrepancy between analytical predictions thus obtained and Molecular Dynamics simulation results for both conservative and dissipative systems.

As the width of the front $\epsilon$ is small compared to the curvature radius of the interface, the flow within can be assumed to be almost one-dimensional. Then, for every field locally defined as $\Psi(r,t)$, we reduce hydrodynamic equations \eqref{eq:hydro} to their 1D expression, and put them under the flux-difference form
\beq \dt \Psi(r,t) + \dr J(r,t) = 0 \eeq
with $J(r,t)$ the corresponding flux.
This form allows us to integrate them between the front boundaries $r_1(t)$ and $r_2(t)$,
\beq \int_{r_1(t)}^{r_2(t)} \dt \Psi(r,t) =\dt \int_{r_1(t)}^{r_2(t)}  \Psi(r,t) +\dot r_1(t)\, \Psi(r_1,t) - \dot r_2(t) \,\Psi(r_2,t)\eeq
which can be simplified since, by assumption, $r_1(t)=R(t)$, $r_2(t)=R(t)+\epsilon$ and $\dot r_1(t)=\dot r_2(t)=R(t)$. The integral term on the left will be proportional to $\epsilon$ and can be expected to vanish compared to the difference term, and we are left with
\beq\left[\,J(r,t) - \dot R(t) \Psi(r,t)\, \right]_{r_1}^{r_2}=0\eeq
where the three notations $\left[\Psi(r,t)\right]_{r_1}^{r_2} \equiv \Psi(r_2,t)-\Psi(r_1,t) \equiv \Psi_2(t)-\Psi_1(t) $ are equivalent.

Rearranging all three Eqs.~\eqref{eq:hydro} to take this form, we thus obtain the so-called Rankine-Hugoniot jump conditions:
\beqa & \left[\,n(u-\dot R)\, \right]_{r_1}^{r_2} =0 \nonumber\\
& \left[\,n(u-\dot R)\,u + p\, \right]_{r_1}^{r_2} = 0 \nonumber\\
&\left[\,n(u-\dot R)\left(\dfrac{u^2}{2}+\dfrac{\Theta}{\gamma-1}\right) + u\,p \right]_{r_1}^{r_2} =0\label{eq:RHgen} \eeqa
where $n(u-\dot R)$ gives the number density of particles that cross the front because its velocity differs from that of the flow, i.e. $\dot R \neq u$. When this factor multiplies a quantity, it represents the ``geometric'' flux of that quantity,  purely due to advection by particles crossing the front; the other terms in the brackets represent energy and momentum fluxes due to the action of pressure through the front. In the first equation, the advected quantity is simply mass, taken here equal to unity.

The external gas being at rest, $u_2 = \Theta_2 = 0$ and we finally find
\beqa& n_1 = \left(\dfrac{2}{(\gamma-1)\,\, Z(n_2)} +1\right) n_2\nonumber\\
&u_1=\dot R \left(1- \dfrac{n_2}{n_1} \right)\nonumber\\
& P_1=n_2 \dot R^2 \left(1- \dfrac{n_2}{n_1} \right) \label{eq:RHimp} \eeqa
where the last two relations are conveniently expressed in terms of the compression ratio $n_2/n_1$.
\subsection{Conservative blast and breakdown at high density}
We start by specifying the conditions for the classic TvNS solution.
In the particular elastic case ($\alpha=1$, hence $\Lambda=0$), and with the standard assumption of an isotropic pressure tensor $\bP=p \bI$ (${\bI}$ being the identity matrix), the Euler equations for a perfect fluid are  recovered from Eqs.~\eqref{eq:hydro}. 
Within the blast, a solution can be found with the ansatz that the fields adopt a scaling form: as argued in Sec.~\ref{sub:TvNS}, the location of a point within the structure of the blast is indexed by a single dimensionless variable $\lambda = r/R(t)$ and we may rewrite the fields as follows
\beqa
n(\br,t) = n_0 M(\lambda), \hspace{10pt} \bu(\br,t)=\dfrac{\br}{t}V(\lambda) \nonumber\\ \Theta(\br,t)=\dfrac{r^2}{t^2}T(\lambda), \hspace{10pt}p(\br,t)=n_0\dfrac{r^2}{t^2}P(\lambda) \label{eq:scaling_elastic}
\eeqa
where the prefactor is the dimensionally-appropriate combination of dimensional parameters $r$, $t$ and the initial homogeneous number density $n_0$
(we recall that mass has been taken equal to unity). 
The choice of nondimensionalization will simply change the functional forms for $M$, $V$, $T$ and $P$, which are the dimensionless profiles that fully characterize the self-similar solution.

Let us further denote by $\delta$ the scaling exponent in $R(t)\sim t^\delta$, so that $\dot R =R \delta /t$.
Equations \eqref{eq:hydro} then become ordinary differential equations on the dimensionless profiles,
\beqa & M' (V-\delta) + M(dV + V') = 0, \nonumber \\
& V'(V-\delta) + V(V-1) +  \dfrac {2 P+  P'} M = 0, \nonumber \\
&T'(V-\delta) +2T(V-1) + (\gamma-1)\dfrac { P} M(dV+V') = 0, \label{eq:cons} \eeqa
where we use the notation
\beq \Psi'=\lambda \dfrac{d\Psi}{d\lambda}\label{eq:defprime}. \eeq
The solution to these equations with boundary conditions \eqref{eq:RHimp} is fully determined once we add the constitutive relation for $P$ as a function of other fields; it is known analytically in the dilute limit $\varphi(n_0) \ll 1$, allowing the ideal gas approximation $Z(n)\approx 1$ and $P\approx MT$.

We do not recall the explicit form of the solution for profiles $M$, $V$ and $P$ (or $T$) found in ~\cite{Sedov1959}, which is long and not very informative. 
Still, using these expressions, all other quantities of interest may be derived. For instance, to estimate the energy of the Trinity bomb, Taylor computed the prefactor in the scaling law  
$R(t)= g(\gamma) (t\sqrt{E_0/\rho_0})^\delta $, that depends on the dimensionless parameter $\gamma$,  using the conservation of energy
\beq\int_0^1 d\lambda \, M(\lambda)\left[ \dfrac{V^2(\lambda)}{2} +\dfrac{P(\lambda)}{\gamma-1} \right] = E_0.\eeq
This prefactor $g(\gamma)$ was found to be close to unity, e.g. for tridimensional hard spheres $g(3/2) \approx 1.08$.

\begin{figure}
\centering
\includegraphics*[width=420pt,clip=true]{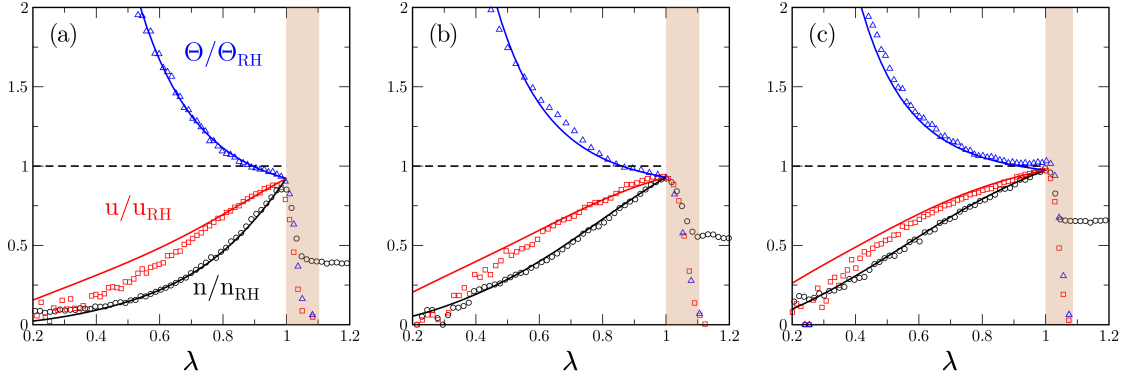}
\caption{Conservative blast at moderate densities. Hydrodynamic profiles rescaled by the theoretical boundary value \eqref{eq:RHimp} at $\lambda=1$ the inner boundary of the shock front (shaded), for density $n/n_{RH}$, velocity $u/u_{RH}$ 
and temperature $\Theta/\Theta_{RH}$. Initial volume fraction: (a) $\varphi_0=0.06$, (b) $\varphi_0=0.20$, (c) $\varphi_0=0.30$. Symbols are simulation results, and solid 
lines are hydrodynamic solutions of Eqs.~\eqref{eq:cons} where we imposed the empirical boundary values. The bulk equations seem to describe empirical profiles very accurately, especially in the outer part of the blast, as the theoretical temperature profile diverges at the center. However, there is a slight discrepancy in the boundary conditions (for instance, $n(1)/n_{RH}< 1$), reflecting the approximations involved in our computation of 
Rankine-Hugoniot conditions. Still, dimensonless profiles for all three fields converge to the same value at $\lambda=1$, meaning that the scaling relations are correct. This is further illustrated in Fig.~\ref{fig:Hyd.Blast.supersonic} below. \label{fig:Hyd.Blast.Hendprof} }
\end{figure}

These profiles have been extensively tested experimentally~\cite{dewey1964air,Edens2004,Moore2005}, and from hydrodynamic simulations~\cite{kamm2000evaluation}. 
However, to the best of our knowledge, they have hitherto not been confronted to microscopic simulations. Furthermore, we provide here
the first approach accounting for higher densities via a more general constitutive relation:  both the boundary conditions and the shape of the profiles are altered by the non-trivial equation of state $p=n\Theta Z(n)$. While not solvable analytically, they can still be integrated numerically with the choice for $Z(n)$ specified in Eq.~\eqref{eq:HendersonZ}. It it thus remarkable to observe in Fig.~\ref{fig:Hyd.Blast.Hendprof} that TvNS-like solutions are confirmed by Molecular Dynamics for low-to-medium densities, except for the boundary values of the fields, whose Rankine-Hugoniot values seem overestimated.  
The slight discrepancy close to the center is expected as the TvNS approach predicts a temperature divergence, which is usually regularized by taking into account 
neglected transport terms such as heat conduction~\cite{abdel1991quasi}.

Even more remarkable is the fact that this extended TvNS regime is limited. As the initial density $n_0$ is increased, the profiles observed in Molecular Dynamics simulations first follow, then diverge from those computed from Eqs.~\ref{eq:cons} and \eqref{eq:HendersonZ} by numerical integration. As we will show, this breakdown hinges on the fact that, in the classic solution, the flow velocity is everywhere subsonic \textit{within the blast} (although supersonic compared to the outside medium) and any perturbation thus quickly diffuses and relaxes. By contrast, high density blasts may exhibit regions of supersonic flow, where the profiles cannot reach their asymptotic self-similar form as perturbations accumulate at the boundary.

\begin{figure}
\centering
\includegraphics[width=300pt]{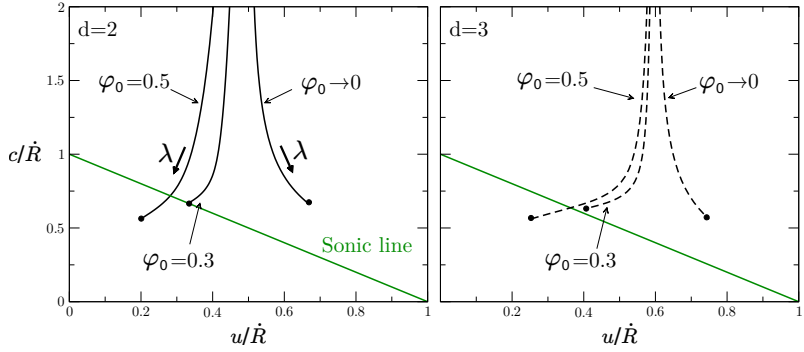}
\caption{Phase portrait $c(u)$ for a conservative blast with finite density: the three curves are parametric representation of the sound velocity as a function of the flow velocity, as obtained by numerical integration of 
Eqs. \eqref{eq:cons}. The dimensionless profiles depend only on $\lambda \in [0,1]$, which is increasing in the direction of the bold arrows. 
Dots correspond to the location of the front $\lambda=1$. We represent solutions in dimension $2$ (left) and $3$  (right) 
for some values of  the initial volume fraction $\varphi_0$. The straight line locates the sonic line $u+c=\lambda \dot R$. 
The critical value of $\varphi_0$ for which the sonic line is reached by the profiles depends on the spatial dimension but stays close to $0.3$. \label{fig:Blast.Hyd.sonicline}}
\end{figure}

\begin{figure}
\centering
\includegraphics*[width=280pt,clip=true]{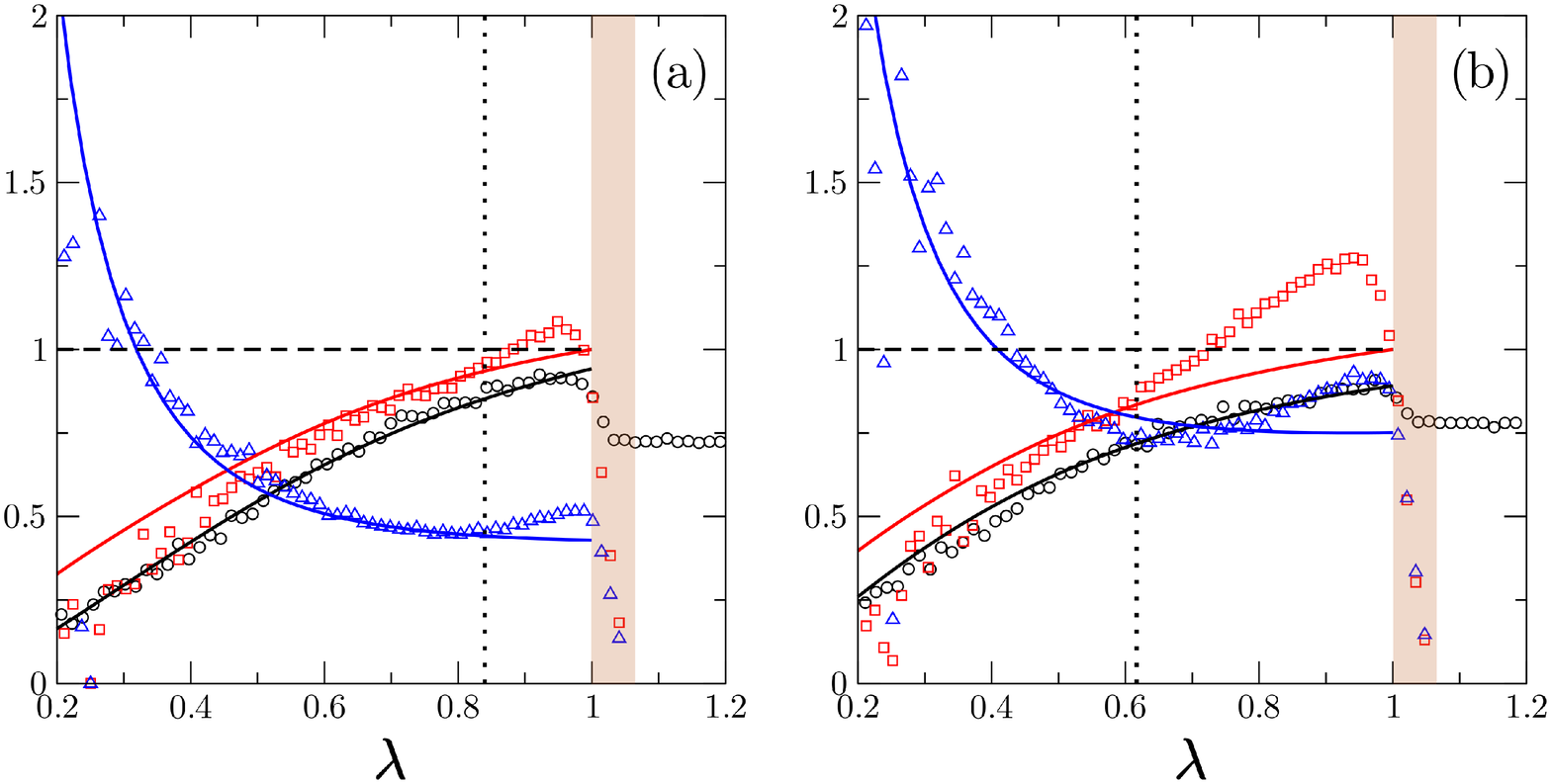}
\includegraphics*[width=120pt,clip=true]{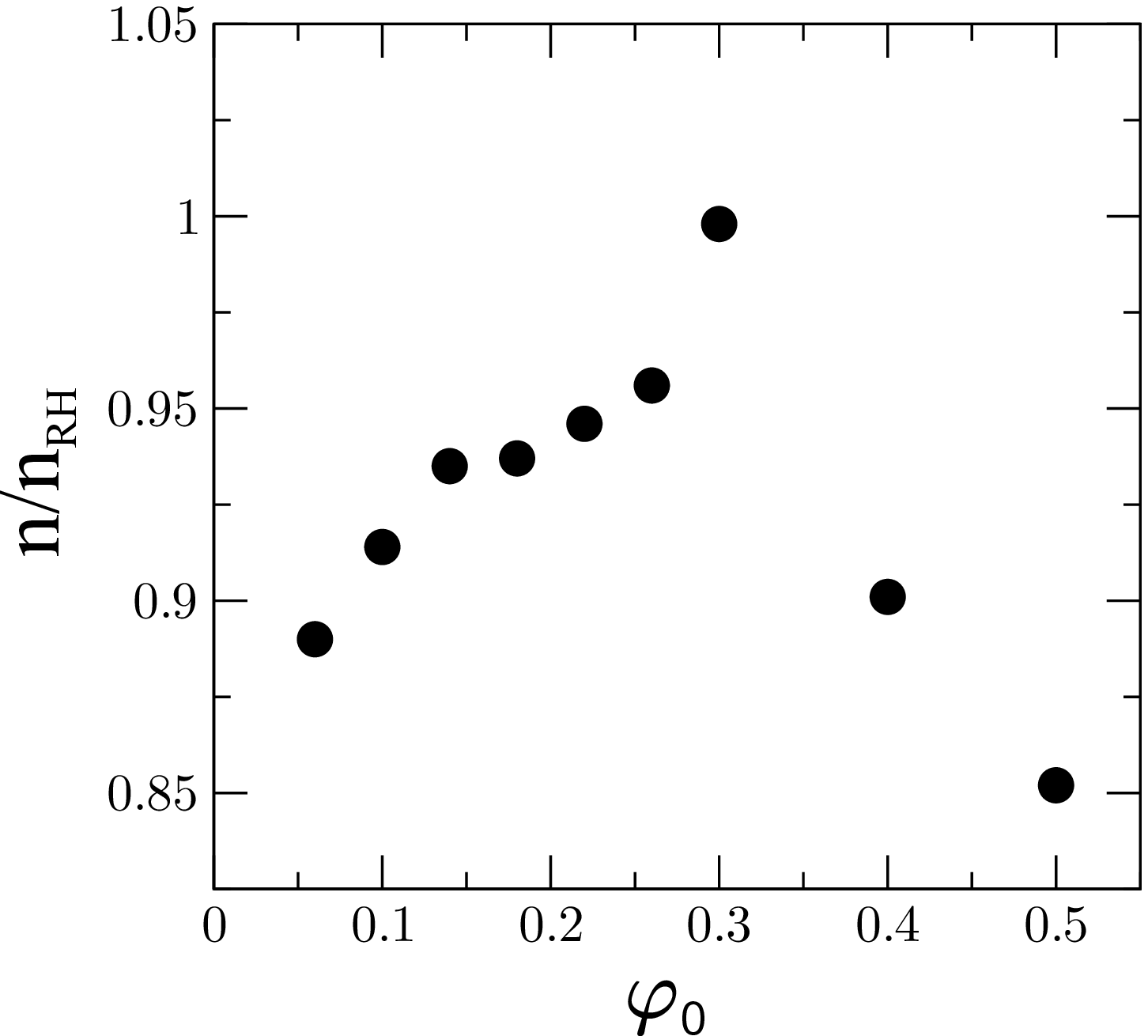}
\caption{Conservative blast at high densities. Density, velocity and temperature profiles (see details in Fig.~\ref{fig:Hyd.Blast.Hendprof}) in two cases where phase velocity is supersonic immediately behind the shock: (a) 
$\varphi_0=0.40$  and (b) $\varphi_0=0.50$. By contrast with the case of entirely subsonic solutions $\varphi_0\leq 0.30$, the profiles disagree with the asymptotic hydrodynamic 
solution (solid lines) and do not converge toward boundary values that are compatible with  each other. The dotted vertical line marks the abscissa where theoretical profiles 
cross the sonic line  $u(r,t)+c(r,t)= \lambda \dot R(t)$, which does indeed appear to be the point where the velocity and temperature profiles deviate from model predictions, with temperature becoming non-monotonic. the rightmost panel, giving the ratio between the measured and theoretical compression at the boundary $n(1)/n_{RH}$ as a function 
of $\varphi_0$, with the surprising effect that best agreement is found for a large value of $\varphi_0 \approx 0.3$. This may be understood as resulting from the fact that the compression 
through the shock is smaller for larger initial volume fractions, thus limiting the importance of higher order effects.\label{fig:Hyd.Blast.supersonic} }
\end{figure}

To discuss this phenomenon, we must define the sound velocity as
\beq c(r,t)=\sqrt{\gamma P(r,t)/n(r,t)}\eeq
which appears naturally by considering the propagation of small compression waves around an homogeneous initial state, and is thus expected to have local validity. Since we have assumed an adiabatic flow without heat waves, 
and there is no transverse velocity field, any perturbation should be in the form of acoustic waves. Thus, the maximal and minimal velocity of an energy-carrying perturbation are typically  $u(r,t) + c(r,t)$  and $u(r,t)-c(r,t)$, depending on whether it is carried by the flow or going against it. By contrast, the phase velocity of the blast is $\lambda \dot R$, the geometric speed of a point at a fixed value of $\lambda=r/R(t)$. If this phase velocity exceeds the maximal, some regions within the blast become isolated from others. This is usually checked by representing profiles $c(r,t)$ and $u(r,t)$  on the parametric graph $c(u)$ (see  Figure~\ref{fig:Blast.Hyd.sonicline}) where the \textit{sonic line} is defined by  
\beq u(r,t) + c(r,t) = \lambda \dot R(t).\eeq
If the profile $c(u)$ stays above this line, perturbations are always faster than the phase velocity; this is the case of the dilute TvNS solution, but not of profiles found for high densities. 
Any singularity in the flow (that does not involve a shock and thus some compression) must lie on the sonic line -- or any other characteristic line corresponding to another relevant physical 
speed being equal to the phase velocity, i.e. $u = \lambda \dot R$ and $u-c = \lambda \dot R$~\cite{Kushnir2010}. We find indeed that this point separates the region where the dense TvNS solution describes numerical 
profiles correctly from that where it fails. This illustrated in Fig. \ref{fig:Hyd.Blast.supersonic}, and in particular evidenced by the temperature profile 
(upper curve), which becomes non monotonous when plagued with supersonic effects. On the other hand, the density profile is still in good agreement with its TvNS counterpart.

\subsection{Dissipative blast}
\label{sec:dissip}

The previous detour through conservative blasts demonstrated the importance of the equation of state. In the case of a dissipative blast, due to the accretion into a shell, asymptotic densities within are necessarily very high, and thus the flow must be described as dense fluid. To the best of our knowledge, the single previous attempt at elucidating the full structure of a dissipative blast~\cite{Bertschinger1986} was committed to an ideal gas description that could not capture the phenomena we observe in simulations, including the instability discussed in Sec.~\ref{sec:instab}.

Instead, we will show that the dissipative coarse-grained equations \eqref{eq:hydro}  admit a solution that closely matches numerical measurements. This solution exhibits an anisotropic pressure tensor acting only along the radial direction $\er$ and of the form 
\begin{equation}
\bp \,=\, n \, \Theta \, Z(n) \, \er \otimes \er 
\label{eq:pression_aniso}
\end{equation}
with $Z(n)$ an adequate choice of constitutive law.

The boundary conditions on these three profiles are set by the fixed width sector, including the front (or
mixing layer) where temperature is created, and the cooling region
where it is dissipated. This sector is not self-similar, but we can make another simplifying assumption to make the equations tractable. If the gradients mostly occur along the normal to the shock boundary, the flow in that region can be approximated as one-dimensional; this approximation improves as this layer's fixed width becomes small compared to its increasing curvature radius. This allows us to simplify
Eqs. \eqref{eq:hydro} by considering them only along the radial direction, and integrate them between  $r=R(t)-x$ and $R(t)$  giving 
\beq
n(x)=n_{\text{rest}} M(x), \hspace{5pt} 
u=\dot R \left[1-\frac{1}{M}\right],\hspace{5pt} 
p=n_{\text{rest}} \dot R \, u .
\label{eq:coolingprofiles}
\eeq
This flux-difference form~\cite{Goldshtein1996} is similar to the one usually employed for the shock front itself, and we recover the Rankine-Hugoniot conditions of Sec.~\ref{sec:RH} by letting $x\to 0$. 
All three fields are parameterized by compression $M(x)$ solving the following equation (which may be integrated
numerically for any choice of $Z(n)$) 
\beqa
%   (M-1) &\left[\left(\dfrac{1}{\gamma-1}+\Lambda_1(\alpha) \right) Z^{-1}+1-\Lambda_1(\alpha)\right]= \nonumber
(M-1) &\left[\frac{d}{2} Z^{-1}+1\right]=
\nonumber \\ 
&\dfrac{M^2}{2} -\omega_0(1-\alpha^2) \int_{0^-}^x dx'
\left(\dfrac{d(M-1)}{2 Z} \right)^{3/2} ,
\label{eq:Blast.Hyd.cooleq}
\eeqa
where the latter term stems from collisional dissipation.
Higher-order transport terms neglected in Eqs.\eqref{eq:hydro} may in fact 
intervene in this intermediary region, which has no growing typical length scale; 
the present simplified analysis proves sufficient for our purposes though.	
 
\begin{figure}[htb]
\centerline{\includegraphics*[width=300pt]{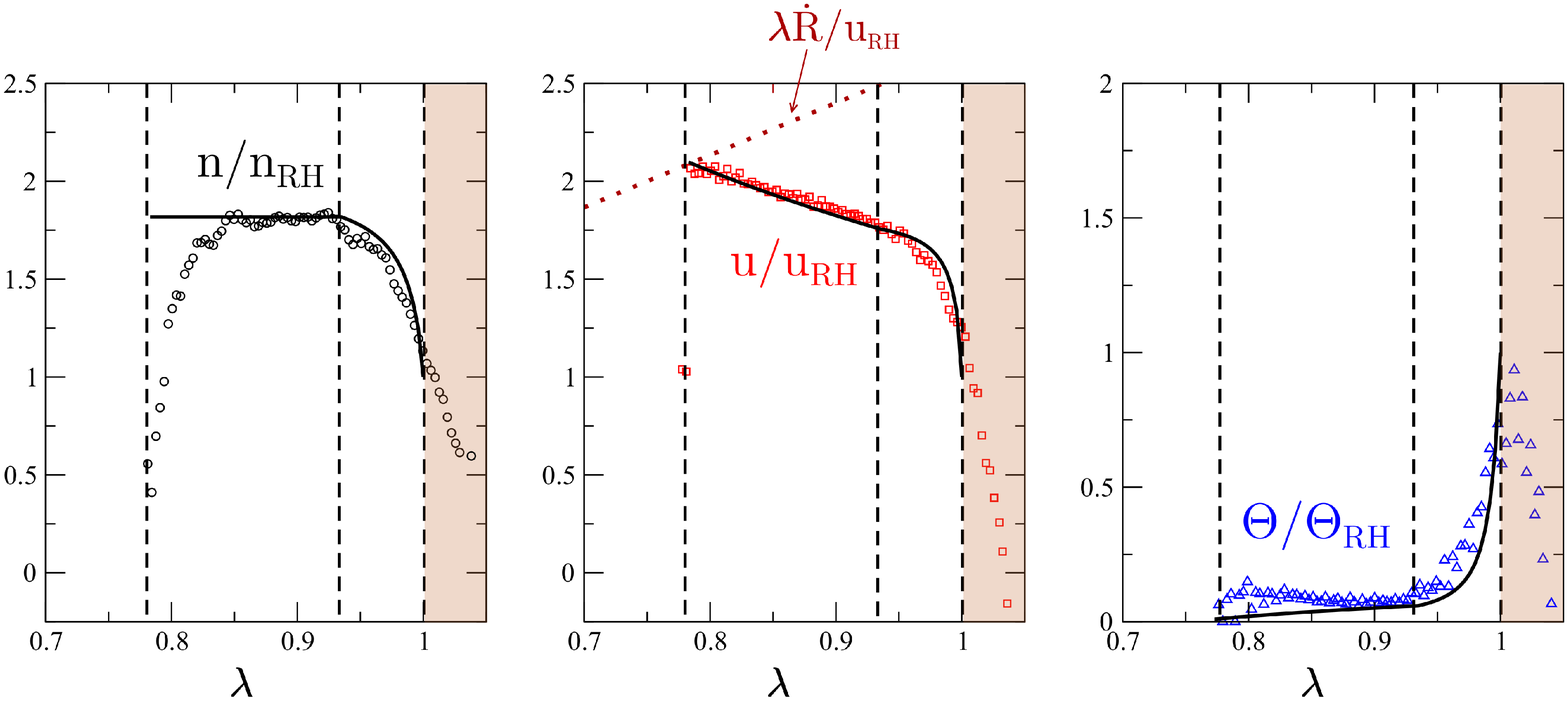}\hspace*{10pt}\includegraphics*[width=110pt]{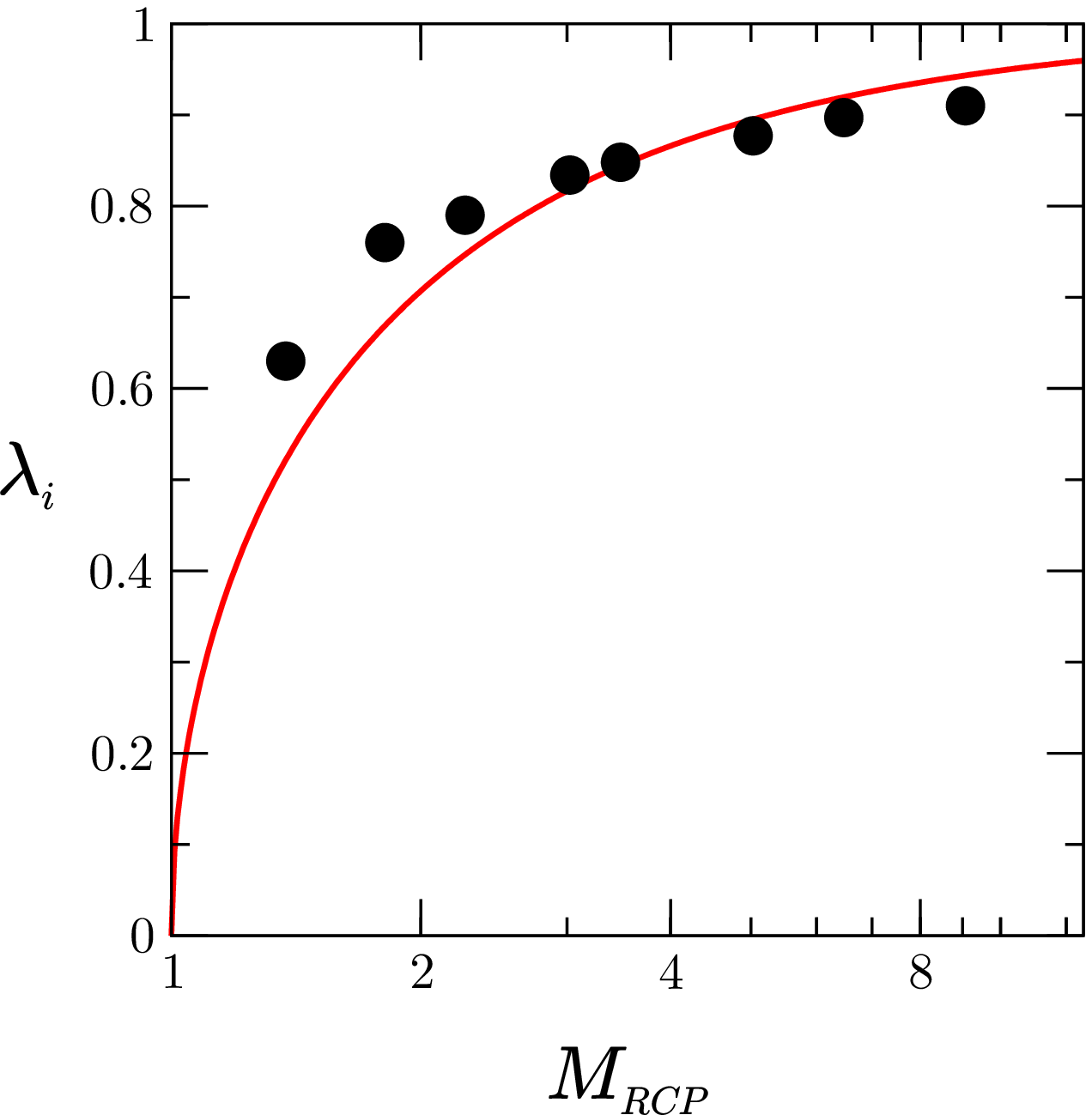} }
\caption{Dissipative blast in spatial dimension $d=2$. Hydrodynamic profiles rescaled by their Rankine-Hugoniot value (Sec.~\ref{sec:RH}) $n(\lambda)/n_{RH}$, $u(\lambda)/u_{RH}$ and $\Theta(\lambda)/\Theta_{RH}$ with 
$\lambda=r/R(t)$. Thick lines are analytic solutions, symbols are measurements from Molecular Dynamics simulations. 
These figures display the spectacular divergence of the solution from the elastic case for any value $\alpha<1$ (here $\alpha=0.8$), with the bulk of the blast dividing into multiple regions. 
Vertical lines represent the inner boundaries of these regions: the front (shaded, the mixing region where some particles are still at rest) which lies above the boundary $\lambda=1$ where fields reach their 
Rankine-Hugoniot value and temperature is maximal; the cooling region between $\lambda=1$ and $\lambda=\lambda_c$ defined as the point where density reaches its Random Close Packing value in the calculated profiles; and the cold gas down to 
$\lambda=\lambda_i$ where pressure vanishes and the flow velocity equals the phase velocity. The rightmost panel compares the approximation \eqref{eq:lami_approx} for $\lambda_i$  as a function of the compression ratio $M_{RCP}=n_{RCP}/n_0$ (solid line) to various late-time measurements in simulations (dots). The approximation ignores the width of the cooling layer, which is asymptotically negligible. Even so, we find a reasonable estimate $1-\lambda_i \approx 0.15$ for the width of the cold region in the profiles shown here, with $M_{RCP}=\approx 3.5$. }
\label{Fig:profiles}
\end{figure}

The fixed-width cooling layer is pushed outward by the expanding cold region, where the absence of temperature allows self-similarity to take hold again, by neglecting 
the main dissipation term in the hydrodynamic equations. We can look for profiles in analogy with Eqs.~\eqref{eq:scaling_elastic}, except for the temperature field: it could previously be used interchangeably with the pressure field, via the constitutive relation, but it is now  vanishingly small while $Z(n)$ diverges due to steric effects. In fact, a simplified analytic solution can be obtained assuming the fluid to be  
incompressible and at random close packing density $n_{\text{rcp}}=n_{\text{rest}}M_{\text{rcp}}$,
a fair approximation corresponding to a volume fraction $\varphi_{\text{rcp}} \approx 0.84$ for $d=2$ or $0.64$ for  $d=3$.  The profiles for velocity and pressure are then
\beqa 
&V(\lambda)= \delta \left( 1 -M_{\text{rcp}}^{-1}\right) \lambda^{-d}, \nonumber \\ 
& P(\lambda)  = \delta^2\,\lambda^2\left( 1 -M_{\text{rcp}}^{-1}\right)\left(M_{\text{rcp}}(
\lambda^d -1) + 1 \right) .
\label{eq:coldprofiles} 
\eeqa
These two solutions can be joined in a piecewise fashion by applying the criterion that the cooling layer stops as soon as the density reaches its random close packing value $n=n_{\text{rcp}}$, at $r_c=R(t) \lambda_c(t)$ (with $\lambda_c(t) \to 1$ since the cooling layer is of fixed width while $R(t)$ increases).  The inner boundary of the cold region, separating it from the empty core, is also rigorously defined: it is given by $\lambda_i$ such that
\beq \dot R \lambda_i = u(\lambda_i),\eeq
meaning that the flow velocity is equal to the phase velocity: this necessarily coincides with pressure vanishing, and with any boundary with an empty region of space, as it means that the velocity of the wave at that point is exactly that of the last particles, i.e. the boundary is defined by these particles moving out and emptying the space behind them.
Assuming that the cooling layer is of negligible width (as is asymptotically true), we then find
\beq \lambda_i \approx \left( 1- M_{RCP}^{-1}\right)^{1/d}. \label{eq:lami_approx}\eeq
Since all other boundary conditions are rigorously computed and have well-defined locations, there is no fitting parameter in the model.
This piecewise solution is seen in Fig. \ref{Fig:profiles} to be in excellent agreement with the profiles measured in Molecular Dynamics simulations, including the locations $\lambda_c$ and $\lambda_i$ of the boundaries between regions.

The boundary $\lambda_i$ is also seen to correspond to the maximal flow velocity in the system, which fits with the intuition that the blast wave moves inertially and is being pushed outward by the innermost particles. The rest are slowed down by the 
dissipative collisions and accrete onto the incoming ``snowplow''. It therefore appears that in the present inelastic case,
radial coherent motion decouples from that taking place in the perpendicular direction (incoherent motion). This is contrast with the elastic situation,
where a single scaling in time does exist for both coherent and incoherent motions. This decoupling lays the ground for the possibility 
of new scaling behaviors, as compared to standard conservative TvNS phenomenology.

\section{Scaling laws}
\label{sec:scal}

\subsection{Conservation of momentum}
\label{sec:momcons}

\begin{figure}[h]
\begin{center}
\includegraphics*[height =85mm,clip=true]{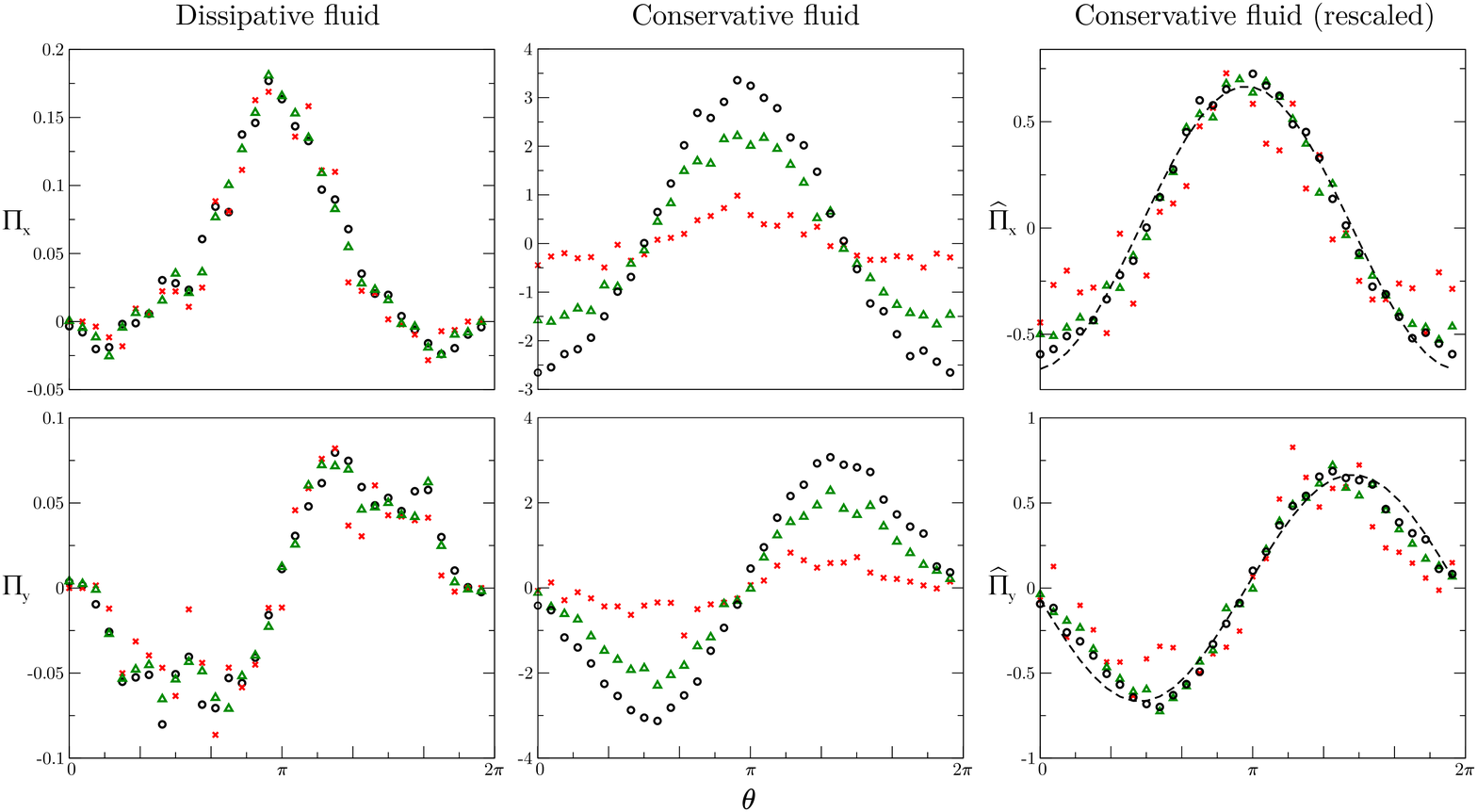}
\caption{Conservation of total momentum per angular sector $\Pi(\theta)$, divided into its $x$- and $y$-component, in a two-dimensional Molecular Dynamics simulation.  
The left panel demonstrates the conservation of $\Pi(\theta)$ over time in the dissipative blast ($\alpha=0.2$), while the central and right panel focus on the energy-conserving blast, 
respectively showing its increasing radial momentum $\Pi(\theta)$ and the same rescaled by the theoretical power-law $\widehat\Pi(\theta)=\Pi(\theta)/t^{d/(d+2)}$. Symbols correspond 
to numerical measurements at successive times $t=1,10,20$ (crosses, triangles and circles respectively), while the dashes give the expected value under the assumption of perfect isotropy, 
$\widehat\Pi_x \propto -\cos\theta$ and $\Pi_y \propto -\sin\theta$. \label{fig:pconserv}}
\end{center}
\end{figure}

Previous works \cite{Jabeen2010} have already confirmed that Oort's proposed asymptotic regime for dissipative blasts, the Momentum-Conserving Snowplow \cite{Oort1951} based on the global conservation of \textit{radial} momentum, is observed in simulations inelastic hard spheres, 
as we show in Fig.~\ref{fig:pconserv}. However, no rigorous justification for that conservation law has been proposed, and evidence in experiments on granular flows is unclear~\cite{Boudet2009,boudet2013unstable}. Using our hydrodynamic model, we can now investigate the conditions under which the conservation of radial momentum can arise in a granular flow, while it is absent from most types of blasts, including the  classical TvNS solution.

Let us first recall the equation \eqref{eq:uhydro} for the local conservation of momentum, and write it in the case of isotropic pressure $\bp= p \bI$ as in the TvNS solution
\beq   \dt (nu) + \divg \left(nu^2 \right)+\dr p =0. \label{eq:Blast.Hyd.Piso}\eeq
Integrating over an angular sector of width $d\theta$ around angle $\theta$ (or likewise with solid angles if $d>2$), we find
\beqa \dt \Pi(\theta,t) = \dt \int_{0}^{\infty} nu \,r^{d-1}dr = \left[nu^2 + p\right]_0^{\infty}-(d-1) \int_{0}^{\infty} p r^{d-2}dr  \eeqa
where we dropped the  $(r,\theta,t)$ dependency. Velocity and pressure are both zero in the gas at rest, and density is vanishingly small at the center of the blast (even in the TvNS solution), so that
\beq \dt \Pi(\theta,t) = p(0,\theta,t) -(d-1) \int_{0}^{\infty} p(r,\theta,t) r^{d-2}dr. \label{eq:Blast.Hyd.Pcons}\eeq
This formula adequately describes a range of possible phenomena. If the central pressure dominates the expansion, the total momentum per angular sector $\Pi(\theta,t)$ increases with time, 
as observed for the conservative blast in the central panel of Fig.~\ref{fig:pconserv}. On the other hand, in a hollow blast solution with no central pressure $p(0,\theta,t)=0$, we find that  $\Pi(\theta,t)$  decreases with time, as observed in certain conservative blasts which are forced to adopt a hollow structure -- either for gases with $\gamma \gtrsim 1$, or when the density of the external medium decreases strongly with radial distance $r$ \cite{Sanz2010}. Orthoradial momentum transfers occur within the shell itself that cause radial momentum to decrease, and thus slow down the expansion, leading to scaling exponents $\delta < 1/(d+1)$.

The exact conservation of radial momentum that is at the core of the MCS solution is only found if orthoradial transfers cannot take place. In extremely dilute fluids such as found in astrophysics, 
the usual argument is that the shell will be so thin that we may neglect any transfers happening within. However, the granular blast has a non-negligible thickness, and still exhibits the scaling regime. Orthoradial transfers vanish identically if the pressure is not isotropic but purely radial, i.e. $\bP = p \er \otimes \er$, in which case the total momentum per angular sector can change only due to central pressure
\beq \dt \Pi(\theta,t) = p(0,\theta,t)\label{eq:Blast.Hyd.Paniso}\eeq
and is therefore conserved in a hollow blast. A strong anisotropy of pressure is in fact commonly observed in granular flows~\cite{barrat2002molecular} and it is expected to appear here due to the existence of a privileged direction. In the Supplementary Material, we discuss the consequences of relaxing this assumption.

The momentum conservation principle that controls the similarity regime of the dissipative blast can thus only be understood in the light of microscopic insight, whereas continuum approaches usually posit isotropic pressure by default and face a contradiction.

\subsection{Intermediate regimes}
\label{sub:regimes}
We may then consider the succession of intermediate regimes that govern the evolution of the system before the MCS phase. A scaling regime is characterized by a value of exponent $\delta$ in the scaling law $R(t)\sim t^\delta$, from which all other scaling exponents can easily be derived by dimensional analysis.
As explained above, both pressure at the center and orthoradial momentum transfers must vanish to enter the radial momentum-conserving regime. 

However, let us assume that orthoradial exchanges in the shell vanish before the central pressure caused by a small number of comparatively energetic particles that populate the hollow core, and have yet to accrete into the shell. Thus, the right-hand term is non-zero in Eq.~\eqref{eq:Blast.Hyd.Paniso}. Further assuming that the very dilute gas in the cavity obeys the law of adiabatic expansion (density is low enough for dissipation and other effects to be negligible), and given that the volume of the cavity is proportional to $R^d$, we have $P (R^d)^\gamma = \text{constant}$.  An intermediate regime, known as the 
Pressure-Driven Snowplow (PDS)~\cite{Cioffi1988}, can therefore arise with exponent $\delta = 2/(d\gamma+2) \geq 1/(d+1)$. It is self-similar of the second kind, varying with 
the adiabatic index $\gamma$ which is a nondimensional microscopic parameter of the dynamics. 

\begin{figure}[htb]
\centerline{\includegraphics*[width=6.5cm,clip=true]{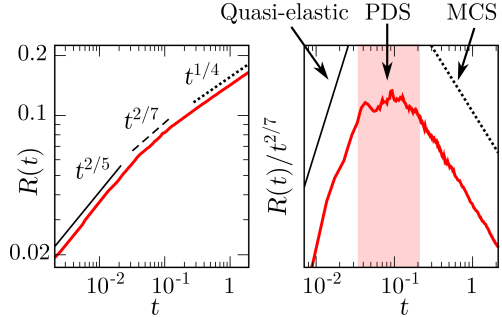}}
\caption{Scaling of the radius $R(t)$ in a three dimensional simulation ($d=3$). There are three successive regimes:  quasi-elastic $t^{2/(d+2)}$ (solid line) before dissipation 
becomes significant, then Pressure-Driven Snowplow (PDS, dashes) $t^{2/(d\gamma+2)}$, and Momentum-Conserving Snowplow (MCS, dots) $t^{1/(d+1)}$. Left: Raw data.  
Right: Data rescaled by the PDS law, to evidence that regime over one decade (shaded region). } 
\label{Fig:scaling_laws}
\end{figure}

This derivation finds its validation as Fig.~\ref{Fig:scaling_laws} gives the first evidence from Molecular Dynamics simulations for this succession of regimes in a granular gas. Since our simulated system is comprised of hard spheres, with the standard adiabatic index $\gamma=1+2/d$, the exponent for the PDS regime is $\delta = 2/(d+4)$. As this causes the PDS and MCS exponents to be accidentally identical in spatial dimension $d=2$, the figure instead presents results from simulations in  dimension $d=3$ where $\delta = 2/(d+4)=2/7$  for the PDS regime and $1/(d+1)=1/4$ for the MCS regime.

\subsection{Comparison with experiments}
Experiments on the granular blast \cite{Boudet2009,boudet2013unstable} have not conclusively revealed the existence of the MCS regime, or even any of the intermediate regimes discussed above; instead, authors propose a logarithmic growth, supported by analytical arguments. We now discuss how the previous analysis may provide hints to understand this apparent discrepancy.

Let us first recall the experimental setting. Small beads are flowing down a slope, a massive ball is dropped from above, hits the slope and imparts energy to surrounding beads, and finally rebounds off, exiting the system. The expanding crown of beads excited by the collisions is then observed in the referential of the flow, moving along the slope.
A crucial aspect of these experiments is that the blast starts with a peculiar initial condition: a hollow central region is created as beads are displaced from the contact area of the ball with the slope.

Following the discussion in Sec.~\ref{sec:momcons}, it is thus possible that the expansion of the blast first obeys a power-law with exponent inferior to $1/(d+1)$: before dissipation causes them to vanish, orthoradial momentum transfers may yet persist inside the expanding shell of beads, while central pressure has been eliminated since the blast is created with an inner cavity. As explained before, this causes radial momentum to decrease and leads to slower expansion than in the MCS regime.
Another possibility is that the similarity regime is simply never observed, as the external medium has finite temperature, and strong shock conditions do not hold for long,  despite the fact that the shock is created with high Mach number $M\approx 10$~\cite{boudet2013unstable}.

Finally, the logarithmic regime proposed by the authors of \cite{Boudet2009} deserves further discussion: it is indeed possible as a transient regime before similarity takes hold. It is not fully self-similar, but driven by energy dissipation in the travelling wave-like cooling layer that immediately follows the shock front (see Sec.~\ref{sec:dissip}), making no assumption on the form of hydrodynamic fields in the innermost regions, except that they exert negligible outward pressure on the cooling layer. As discussed in a previous work~\cite{blast1d}, a very similar phenomenon is seen in a one-dimensional system, where the shock front and cooling region at first reduce to a single leading particle, which decelerates geometrically with collisions and thus propagates logarithmically. It is only when following particles catch up and push it from behind that the wave evolves toward the eventual scaling regime. In higher dimensions, this regime typically requires the suppression of central pressure to be observed, as it holds only as long as the core cannot push outward on the shell (consistently with an initial hollow core, transiently filling up with particles and energy that drop out of the shell), and it is thus a good candidate to explain experimental findings.

\section{Corrugation instability}
\label{sec:instab}

\begin{figure}
\centering
\hspace*{-10pt}
\includegraphics[width=460pt]{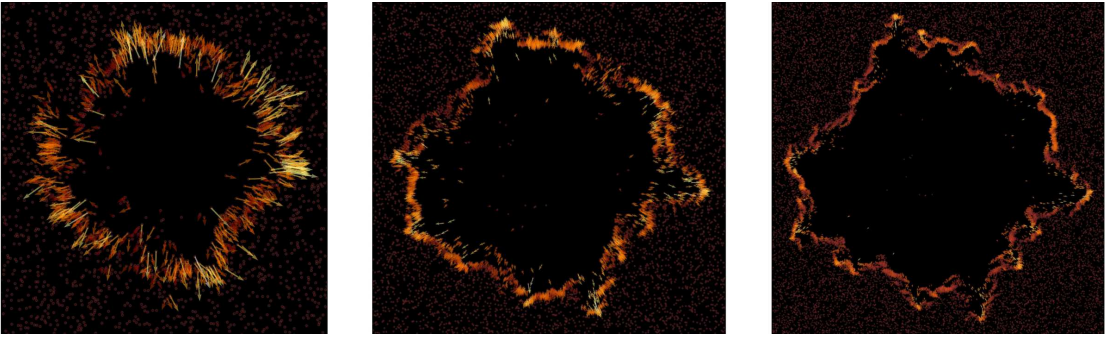}
\caption{Blast particles at times $t=1,8,20,40$ for $\alpha=0.3$ and $\varphi_0=0.1$. All distances are rescaled by $R(t) \sim t^\delta$, and the initial excitation was distributed over a circular 
region to clearly observe the growth of the instability around the self-similar solution (without confusing it with some initial anisotropy of the blast boundary that would have been preserved 
from the transient phase, as discussed in the main text).\label{fig:Blast.Instab.obnum}}
\end{figure}

Numerical studies of the granular blast \cite{Jabeen2010,Pathak2012,Pathak2012b} have remarked that the shape of the shock boundary is preserved: contrary to the conservative case where central pressure pushes particles outward isotropically, the dissipative blast region does not tend to evolve toward central symmetry with a clear circular (or spherical) boundary. Instead, it retains any anisotropy it had at the start of the self-similar regime, as a result of random collisions during the transient phase, see Fig.~\ref{fig:cover}. Symmetry can be imposed to this shape by distributing the initial excitation over a disk or sphere, or by using particles with low enough inelasticity that the blast first becomes isotropic in a quasi-elastic phase before evolving toward the dissipative asymptotics. 

This peculiar property of marginal stability should not be confused with the previously unobserved phenomenon that we discuss here: whether anisotropic or not, 
the basic self-similar shape of the boundary is later overtaken by a corrugation instability (see Fig. \ref{fig:Blast.Instab.obnum}) that causes the appearance of ripples, themselves growing self-similarly. 
This can be revealed by a linear stability analysis. 

\subsection{Linear stability analysis}
We will focus on the two-dimensional case with central symmetry, as it is most relevant to possible experiments. All hydrodynamic fields must now be written as the sum of their asymptotic self-similar expression, and a small perturbation
\beqa & n(\br,t) = n_0(r,t) + \deln (\br,t) \nonumber \\
& \bu(\br,t)= u_0(r,t) \er + \delu(\br,t) \nonumber \\
& P(\br,t) = P_0(r,t) +\delP(\br,t) \eeqa
The perturbation of the velocity field actually separates into radial $\delur$ and orthoradial $\deluT$ components
\beq \delu = \delur(\br,t) \er + \deluT(\br,t) \eeq
Likewise, we separate the gradient operator into its radial and orthoradial parts
\beq  \grad = \er \dr +  \be_\theta \partial_{\theta} =\er \dr +  \gradT \eeq
We must now specify a functional basis on which to decompose the perturbations. When the equations obeyed by the perturbations have constant or simple coefficients, 
it is customary to look for perturbations that are exponential in space and time, so that, for an arbitrary field $\psi$, the perturbation and its derivatives exhibit similar scaling
\beq \dt \delta\psi \sim u \dr \delta\psi \sim \delta\psi.\eeq
Here however, the equations involve unperturbed fields which are self-similar.
Therefore, $\dr=R^{-1} \partial_{\lambda}$ and $u_0 \sim R/t$ meaning that

\beq u_0\dr \delta \psi\sim \dfrac{\delta \psi}{t}.\eeq
On the other hand, the perturbations are seen to cause global transport of mass, momentum or energy, and this requires the local temporal 
derivatives and transport terms to be of the same order in time (else they would decouple asymptotically). 
This requires that the perturbations exhibit a power-law scaling with time, so that  $\dt \psi \sim \psi/t$ also holds. Finally, a natural basis for the angular dependence is that of sinusoids. Hence, we choose to write the elementary perturbation as
\beq \dfrac{ \delta \psi(\br,t)}{\psi_0(\br,t)} = t^s  \delta\Psi(\lambda)\, \cos(k\theta)\eeq 
Likewise, the perturbed radius is given by
\beq R(\theta,t) = R_0(t) + \delR(\theta,t), \qquad \qquad \delR(\theta,t) =R_0(t)\,  t^s \cos(k\theta). \label{eq:Rperturb}\eeq
With this expression, exponent $s$ reflects the stability of the base solution to a perturbation with angular number $k$: if $s>0$, that perturbation grows self-similarly faster than the perturbed solution, if $s=0$ it is marginally stable, and if $s<0$ it disappears asymptotically. Such perturbations have no characteristic timescale; by contrast, exponential 
perturbations would be resolved locally since the time derivatives would asymptotically dominate any transport term.
Under the above provisos, we obtain the evolution law
\beqa
&\left(\begin{array}{ccc}1 & 0 & 0 \\ V-\delta & 0 &M^{-1} \\ 0 & V-\delta & 0 \end{array}\right)\dfrac{d}{d\ln \lambda} \left(\begin{array}{c}\delVr \\ \delVT \\ \delP \end{array}\right) = \nonumber \\  
&\qquad-\left(\begin{array}{ccc} d & -k^2 & 0\\ s-1+2V + V' & 0 &(d+1)M^{-1} \\ 0 & s-1 + V & M^{-1}\end{array}\right) \left(\begin{array}{c}\delVr \\ \delVT \\ \delP \end{array}\right)
 \label{eq:Blast.Instab.matrix}\eeqa
where we recall from Eq.\eqref{eq:defprime} that  $\Psi'=d\Psi/d\ln\lambda$  for any field $\Psi(\lambda)$.
In Eq. \eqref{eq:Blast.Instab.matrix}, the angular number $k$ is chosen, while 
the growth exponent $s$ is unknown, and to be determined from a self-consistent procedure
detailed below.

\subsection{Boundary conditions}

The inner boundary condition is easily defined: it is located at the contact of the dense shell and the empty core, where total pressure must vanish. As we see that the inner and outer interfaces exhibit the same corrugation, the former sits at $R_i(\theta,t) = \lambda_i R(\theta,t)$. We recall that by approximation \eqref{eq:lami_approx}, $ \lambda_i \approx \left( 1- M_{RCP}^{-1}\right)^{1/d}$
where $M_{RCP}= n_{RCP}/n_0$ is the maximal random compression. Thus, the inner boundary condition is given by
\beq p(\lambda_i R) = p_0(\lambda_i R_0) + \lambda_i \delR \;\dr p_0(\lambda_i R_0) + \delP(\lambda_i R) = 0\eeq
where $p_0(\lambda_i R_0)=0$, and
\beq \delP(R_i) = - \lambda_i \delR \dr P_0(\lambda_i R_0). \label{eq:condelP} \eeq
This condition translates for dimensionless variables to
\beq \lim_{\lambda \to \lambda_i} P(\lambda) \delwP(\lambda) = -  P'(\lambda_i) = -d \delta^2 M_{RCP}\label{eq:condelwP} \eeq

The outer boundary conditions will be situated at the limit of the cooling region. The unperturbed fields satisfy Rankine-Hugoniot conditions \eqref{eq:RHimp} at the unperturbed position  $R_0(t)$, and similar conditions must apply to the perturbed fields at the new position $R(\theta,t)$ defined in Eq.~\eqref{eq:Rperturb}.
At this point, the value of the pressure field is given (to first order in $\delR$ and $\delP$) by
\beq p(R,\theta,t) \;= \;p_0(R) + \delP(R,\theta)\; \approx \;p_0(R_0) + \delR \;\dr p_0(R_0) + \delP(R,\theta) \eeq
and similarly for other fields.
The normal to the interface is not the radial unit vector $\er$ anymore, but a distinct vector $\be_n$.
Assuming a small perturbation $\delR \ll R$, we may write $\be_n$ and the tangent vector $\be_t$ as
\beqa &\be_n \approx  \be_r - \left(\partial_y  \delR\right)     \be_y \nonumber\\
 &\be_t \approx \left(\partial_y  \delR\right)   \be_r  +   \be_y . \eeqa
Thus, the normal velocity appearing in Rankine-Hugoniot conditions is now 
\beq\bu(R).\be_n  u_0(R_0) +\delR \;\dr u_0(R_0) + \delur(R)\eeq
and the tangential velocity must vanish
\beq \bu(R).\be_t \approx  \left(\partial_y  \delR\right) u_0(R_0)  - \deluT(R) =0. \eeq
Finally, 
\beqa &\delM(1) = - M'(1) 
& &\delVr(1) = \dfrac{s}{\delta}V(1) - V'(1) \nonumber\\
& \delVT(1) = -V(1) & &
\delwP(1)=\dfrac{2s}{\delta} P(1) -  P'(1)
\label{Blast.Instab.Stablin.RHpert} \eeqa
where $M$, $V$ and $P$ are the dimensionless fields defined in \eqref{eq:coldprofiles} and their derivatives on the boundary can be computed using the hydrodynamic equations. Since we use the incompressible approximation, the first equation simplifies to $\delM(1)=0$.

\subsection{Shooting method}
Profiles for the eigenmodes of the perturbation can be obtained by integrating Eqs. \eqref{eq:Blast.Instab.matrix} 
with boundary conditions \eqref{Blast.Instab.Stablin.RHpert}. These contain two parameters: $k$ the wave number, and 
$s$ the growth exponent, which appears both in the bulk equations and in the boundary conditions. For any value of $k$, 
there is only one value of $s(k)$ that allows the profiles to satisfy both outer and inner boundary conditions; 
however, analytical integration is not possible, therefore we must use a shooting algorithm: for every sampled $k$, 
we test values of $s$ by using them to integrate the profiles numerically from the outer boundary inward, and we 
select $s(k)$ which minimizes the distance between the measured profiles at  the inner boundary and their theoretical 
value. This method, proposed in \cite{Kushnir2005} to study instabilities in a different class of blasts, allows to extend linear 
stability analysis in the case of reference solutions which are neither uniform nor stationary, and thus do not easily 
allow for analytical treatment. This approach is however cumbersome and encounters numerous convergence problems, as the 
boundaries are singular points and perturbation profiles are often found to diverge in that limit (see discussion in the Supplementary Material).

Nevertheless, Fig.~\ref{Fig:disp} demonstrates that this method produces conclusive results. The dispersion relation $s(k)$ 
is found for each of the three eigenmodes with wave number $k$ in the set of three equations \eqref{eq:Blast.Instab.matrix}. 
Two of these modes are oscillatory in nature and convergent ($\Re(s) <0$, $\Im(s) \neq 0$) while the last mode reveals a 
non-oscillating instability: $s(k)$ real and $s(k)>0$ for $k>k_c$.

While the shooting algorithm has increasing convergence problems for
high $k$, which do not allow to sample arbitrarily high values of the
wave number, the obtained dispersion relation for the leading mode
appears to have a maximum at $s\approx 0.3$, or perhaps even a plateau
that might extend indefinitely. The ordinate of this plateau does not
depend on any of the parameters of the model, e.g $\alpha$ or
$\varphi_0$, as they are entirely absent from the stability analysis
above. Crucially, this value $s\approx 0.3$ is precisely the measured
exponent for the relative growth of the corrugation amplitude in
simulations, which is also independent on these parameters (see Fig. \ref{Fig:disp}). 

Let  us now discuss the peculiarities of this phenomenon, and in what it
  differs from previously reported classes of instabilities. First, the growth of the
  corrugation is itself self-similar even for short times. This distinguishes it from
  the best-known Rayleigh-Taylor type, characterized by an initial exponential
  growth, in which self-similarity can occur only for late times
  and under certain specific conditions~\cite{cook2004mixing}. The
  power law structure suggests that the instability takes place in the
  cold gas region, and, as our analysis shows, 
  hinges on the layered structure of the solution. For that reason, the details of the microscopic
  parameters like the value of inelasticity or the initial density
  of the gas do not affect the behavior of the corrugation, giving a
   universal fingerprint to this phenomenon. The vanishing
  pressure at the center is a distinctive characteristic and a necessary
  condition for the instability to take place. If a finite pressure is
  imposed at the boundary, as in a piston-like setup, the
  instability completely changes in nature, losing its self-similar
  structure and showing the presence of convective rolls~\cite{sirmas2015evolution}. On
  the contrary, the vanishing pressure ensures the conservation of
  momentum per angular sector, which is maintained even after the
  corrugation instability appears.
%  , as shown in  Figure~\ref{sec:momcons}. 
This leads to the following
  interpretation: density fluctuations in the gas
  produce a slowing down of some points of the front and a consequent
  discrepancy in the propagation velocity of neighboring regions, producing the
  effect of slosh dynamics between the cold gas and the particles at
  rest. This mechanism is quite different from the one proposed
  in the blast waves propagating in uniform gases: in that case, a net
  pressure wave along the front of the blast is observed, producing
  oscillations in the hydrodynamics fields. This latter description is
  more consistent with a dispersion relation with an imaginary
  contribution, instead of the pure real value of the exponent $s$
  observed in our case.

\begin{figure}[h]
\includegraphics*[height=120pt]{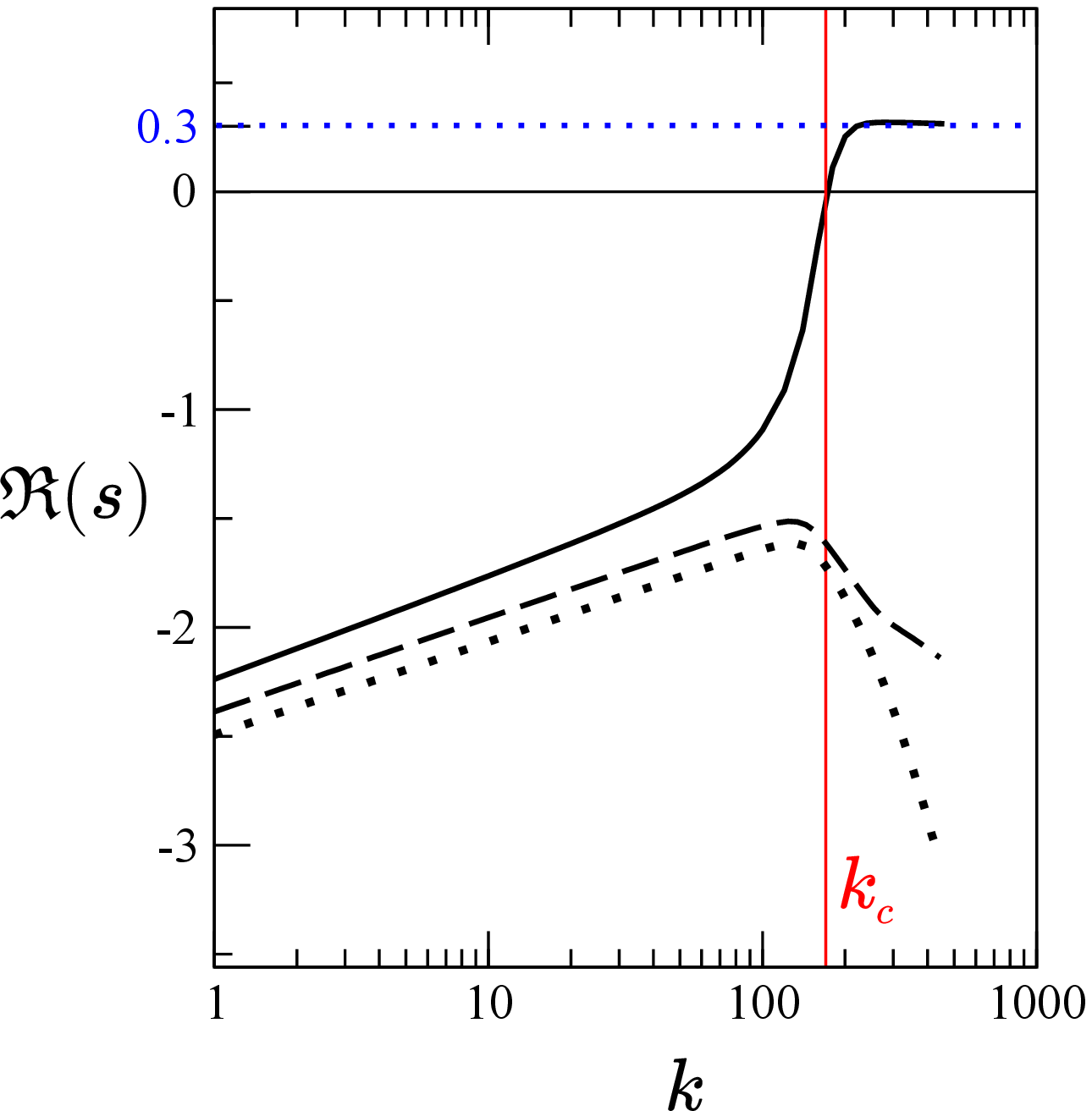}\hspace*{3pt}\includegraphics*[height=120pt]{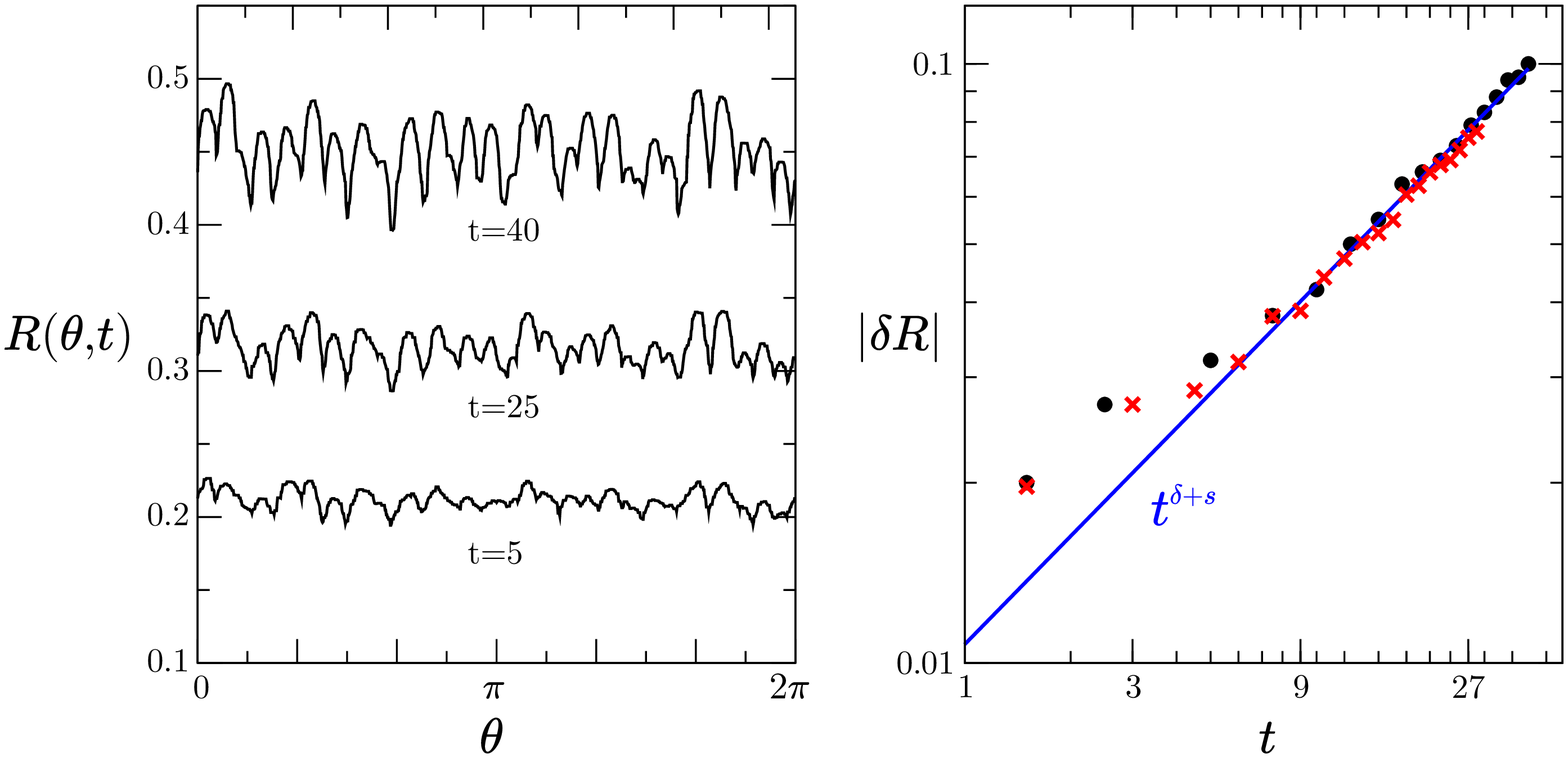}
\caption{
%Top left:\XXXX Cartoon of the instability mechanism.
The corrugation instability in a dissipative blast.
Left: real part of the dispersion relation $\Re s(k)$ for all three modes. The unstable mode crosses the origin at $k_c\approx 160$ and has a plateau at $s\approx 0.3$. It is also strictly real, while the two vanishing modes have constant imaginary components (not shown). Center: plot of the corrugation growth $R(\theta,t)$ for different  times. 
Right: numerical verification of the exponent $s\simeq 0.3$ for values of $\alpha$ equal to $0.3$ (circles) 
and $0.8$ (crosses).}
%mechanism + illustration of
%  the corrugation growth + measure of growth exponent.}
\label{Fig:disp}
\end{figure}

\section{Conclusion}
We have established a bridge between the coarse-grained view of strong shocks and a microscopic level of 
description. While we have put the emphasis on dissipative media, our analysis led us to revisit the more traditional 
problem of Taylor-von Neumann-Sedov blasts in conservative fluids. It appears  that in both cases, the dynamics of the blast exhibits self-similarity of the 
first kind, therefore grounded in conservation laws rather than in microscopic details. Yet, the corresponding scaling laws
are drastically affected by energy dissipation, a key feature being that coherent and incoherent motion
decouple in the dissipative 
case, while they are always coupled, and therefore endowed with similar scaling behavior, when energy is conserved. 
We have discussed the microscopic origin of momentum conservation in different angular
sectors of the shocks, which characterizes the inelastic dynamics and from which 
scaling exponents readily follow, stressing that this conservation law breaks down 
for elastic situations, although binary collisions conserve momentum in all cases.
Furthermore, depending on initial conditions, intermediate nontrivial scaling regimes can be evidenced, 
see e.g. Figure \ref{Fig:scaling_laws}.

Beyond scaling, we have provided an accurate hydrodynamic description
for the coupled density, pressure and kinetic temperature
fields. We have shown that the simple Taylon-von Neumann-Sedov solution can be adapted for dense conservative fluids, and successfully describes a flow simulated at the kinetic level, but breaks down at high densities with the development of supersonic regions within the blast.
On the other hand, shocks in dissipative fluids exhibit a rather complex structure, with an
asymptotic hollow core that has no counterpart in conservative blasts. Its
multi-layered structure is an important ingredient in a successful analytical
description, together with the inertial nature of its
dynamics. Finally, we have shown that the self-similar solution
brought to bear is plagued by a corrugation instability, which
  exhibits some peculiar features. Indeed, the structure of this
  instability is itself similar and does not depend on the microscopic
  parameter of the system, nor on the specific geometry studied
  (similar behaviors are found in planar or three-dimensional geometries, as reported in~\cite{barbierthesis}). 
  Essential for the occurrence of the corrugation are the development of a zero-temperature layer of accreted particles moving coherently, and the
  vanishing pressure in the innermost region of the blast. 

%The description given above cannot be reduced to those already reported
%  in other blast waves even if is probably not peculiar only of the
%  dissipation mechanism introduced here: a deeper understanding of the
%  complete generality of this result is then still an open and
%  challenging issue.  

\section*{Acknowledgments}  
We thank  J. F. Boudet, A. Vilquin, H. Kellay and P. Krapivsky for discussions.
\bibliography{library2}
\setcounter{figure}{0}
\renewcommand{\thefigure}{A\arabic{figure}}

\section*{Appendix A: Numerical methods}
\subsection*{Hard Sphere Molecular Dynamics}
\subsubsection*{Overview}
We consider a $d$-dimensional simulation box of size $L=1$ containing $N_{\text{tot}}$ spherical particles with mass $m$ and 
radius $\sigma$. This defines the volume fraction $\varphi_0=\varphi(n_0)$ occupied by the particles:
\beq \varphi(n_0) = n_0\, \mcl V_d \,\sigma^{d} \label{eq:defphi}\eeq
where   $\mcl V_d$ is the volume of the $d$-dimensional unit sphere and $n_0$ the initial particle density. 
Furthermore, the particles are given a restitution coefficient $\alpha$ such that the kinetic energy dissipated 
in one collision is proportional to $1-\alpha^2$. We simply fix restitution coefficient to a constant value, independent of the velocities of the particles, contrary to other models of granular systems~\cite{Poschel2003}. 
Here, this choice entails no loss of generality, as we show that our results do not depend significantly on the value of $\alpha$, unless it is $1$ (corresponding to elastic particles). However, we must impose a regularization threshold $v_r$ for the relative velocity of the collision partners, under which $\alpha$ for that collision is taken to be $1$, to avoid the problem of inelastic collapse discussed in \cite{Mcnamara1994}.  

The initial configuration is a random hard-sphere distribution over the system computed 
using the pivot method \cite{Krauth2006}, with all particles initially at rest, save for those contained in initial 
radius $r_i$ among which velocities are randomly chosen according to a $d$-dimensional normal distribution, then rescaled so that their total energy is $E_0$.

Unless otherwise specified, all simulations considered in this article were done in dimension $d=2$, 
with $N_{\text{tot}} = 2 \; 10^{5}$, $\varphi_0 = 0.05$, $\alpha = 0.8$ for inelastic particles ($\alpha=1$ for 
elastic particles), unit total initial energy $E_0=1$ and particle mass $m=1$. Whenever computational resources allowed, 
the simulations were run until one of the blast particles had reached the box boundary.

\subsubsection*{Simulation time} Simulation time was made nondimensional so as to become independent of $N_{\text{tot}}$ 
for fixed values of the other parameters, restituting the theoretical expressions in \cite{Visco2008}. 
The unit time $\tau$ is derived from the average collision time in the entire system:
\beq \tau =\sqrt{\dfrac{E_0}{mN_{\text{tot}} }} \dfrac{\varphi_0 \, \chi(\varphi_0)}{\sigma \mcl V_d} \eeq
where $\chi(\varphi_0)$ is the Enskog correction that accounts for increased collision rate at high densities due to the particles  having finite radius. Its expression for $d=2$ takes the following form \cite{Henderson1975} 
\beqa \chi(\varphi)= \dfrac{1- \frac{7}{16}\varphi}{(1-\varphi)^2}. \eeqa

\subsection*{Details on the shooting method}
\begin{figure}
\centering
\includegraphics[width=360pt]{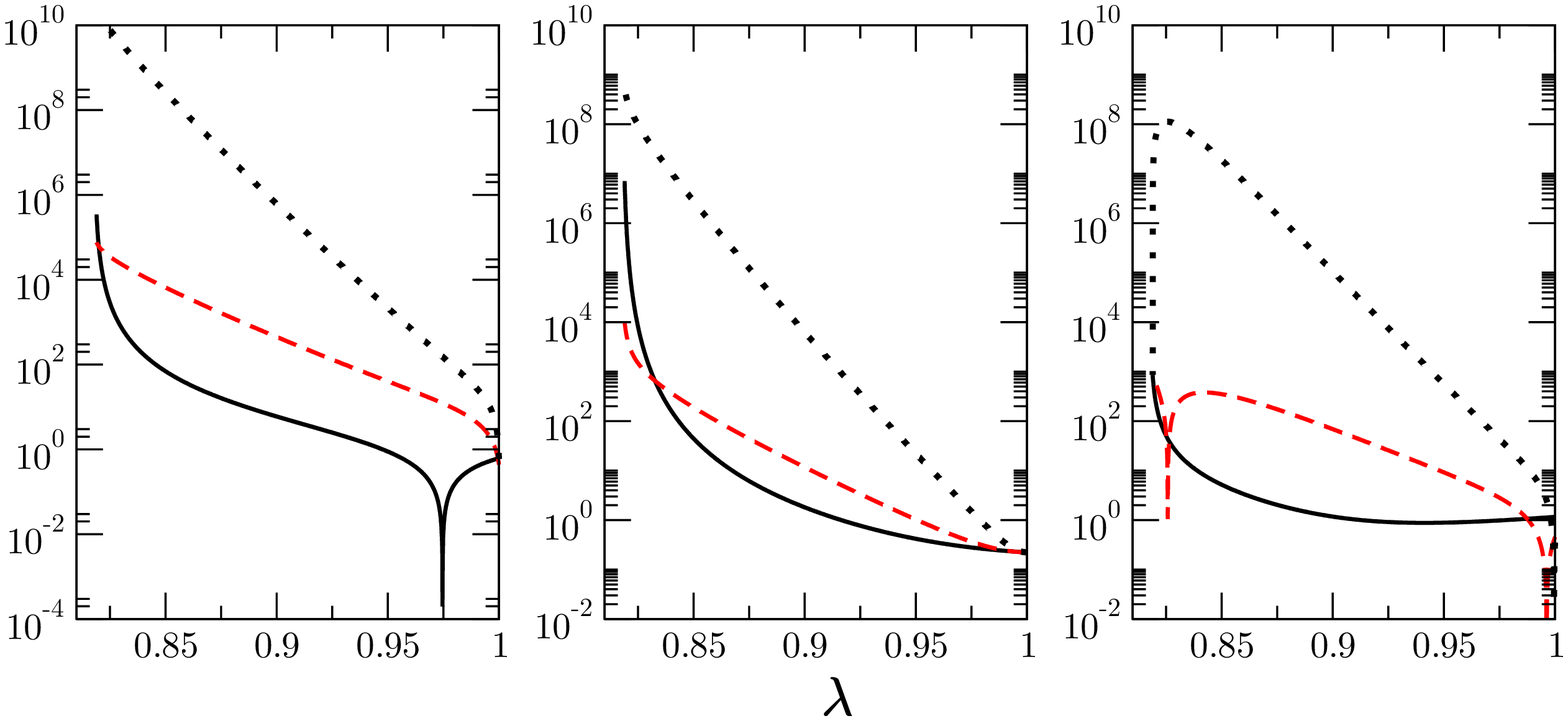}
\caption{Computed perturbation profiles for $\varphi_0=0.3$. From left to right, $|\delVr|$,  $|\delVT|$ and $|\delP|$ on a logarithmic scale for $k=10$ (solid lines), $k=k_c \approx 40$ (dashed lines) and $k=100$ (dotted lines). \label{fig:inelprof}}
\end{figure}

As noted in the main text, the shooting method encounters a divergence. Due to the vanishing pressure at the inner boundary of the blast, perturbation profiles are indeed seen to diverge at that point, as the free interface is infinitely responsive to infinitesimal changes. These profiles are shown in Fig.~\ref{fig:inelprof}. Absolute values have been taken to allow for logarithmic representation, but it is useful to note the signs of the profiles: on the internal boundary $\delVr \to +\infty$ while $\delVT$ and $\delP \to- \infty$. In the unstable cases (dashes and dots), the pressure perturbation $\delP$ changes its sign very close to the internal boundary: it is negative, then positive with large magnitude before returning to $0$. This can be understood as a displacement of the pressure profile, decreasing near the boundary to increase a little further down. This evokes the mechanism of translation by a perturbation discussed by Barenblatt~\cite{Barenblatt1996}. However, in contrast to that classic example, our case involves a temporal scaling $\delP \sim t^{\delta+s}$ that differs from that of the unperturbed self-similar solution.

\begin{figure}[h]
\includegraphics*[height=160pt]{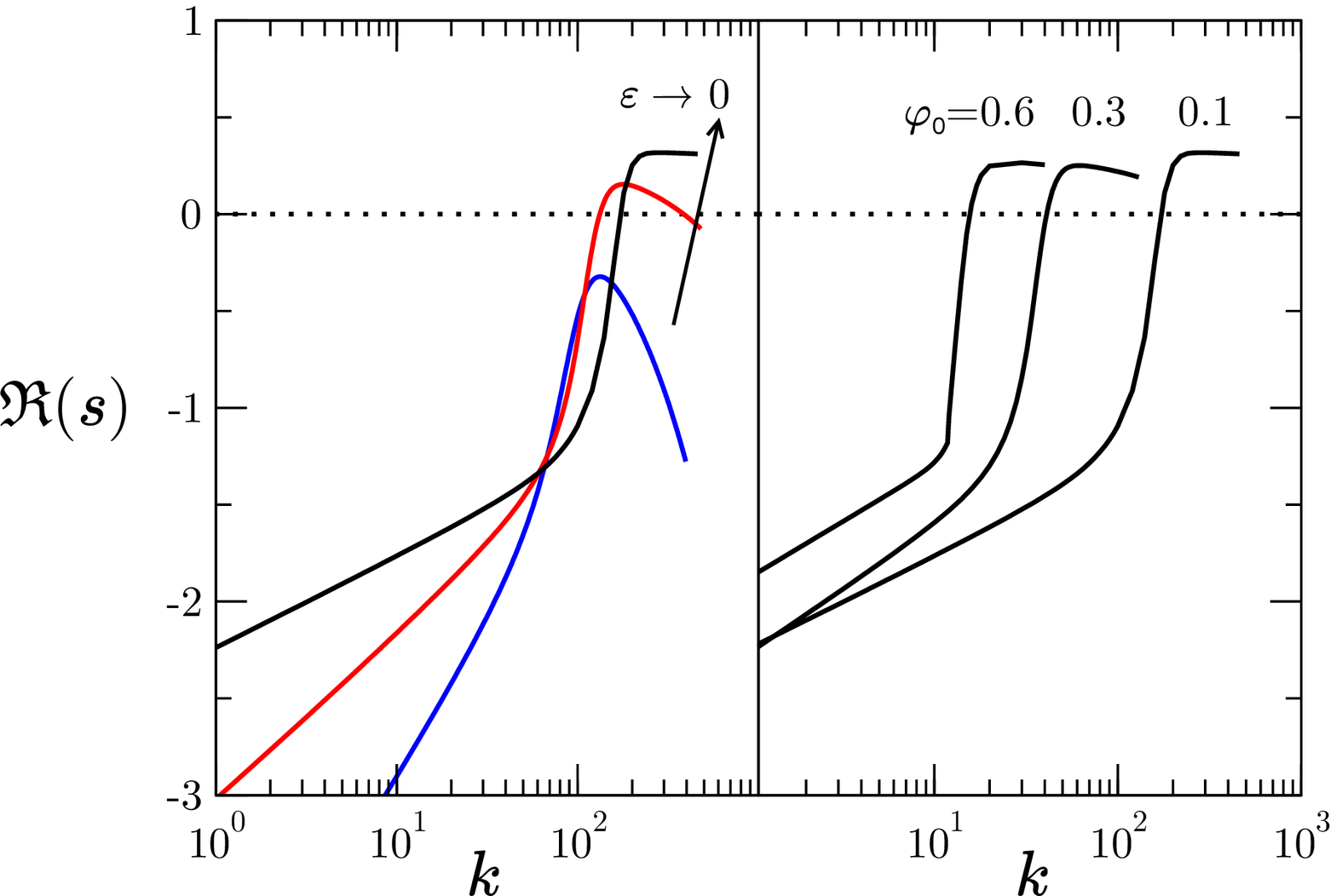}
\includegraphics*[height=120pt]{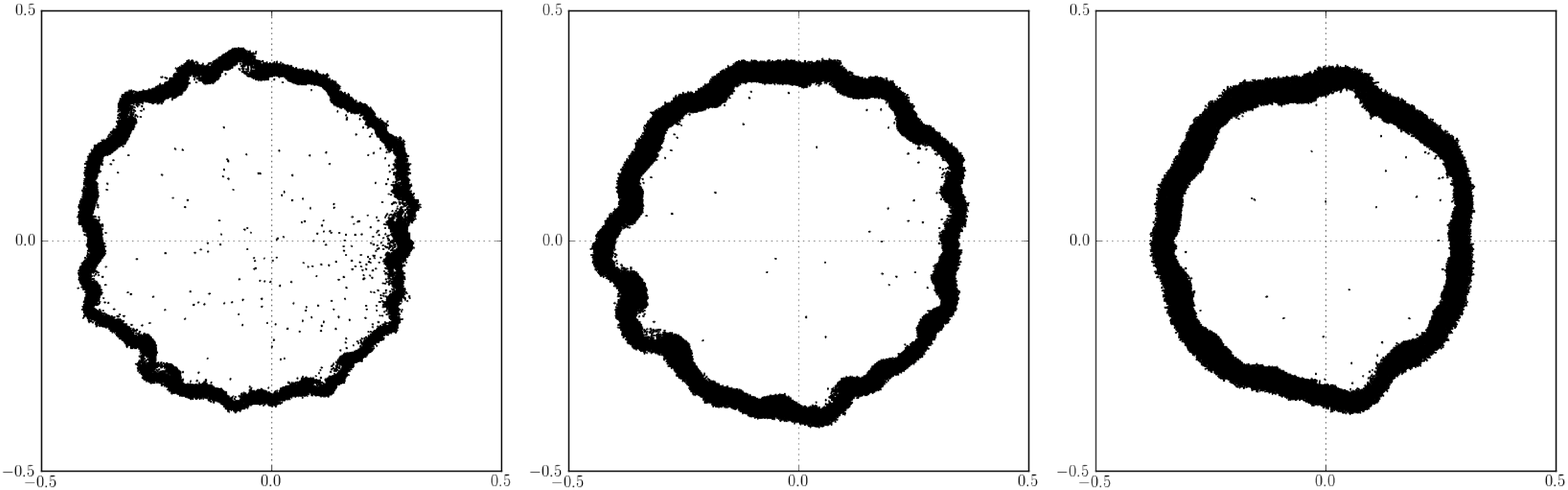}
\caption{Details on the shooting method. Top left: as $\varepsilon \to 0$, the unstable mode converges toward the shape shown in the main text and seems to develop a plateau at $s\approx 0.3$. Top right: Changing the initial volume density $\varphi_0$ does not alter the dispersion relation $s(k)$ significantly, but the onset of instability $k_c$ such that $s(k_c)=0$ is displaced to higher values (higher spatial frequencies) with lower densities. This qualitative behavior is consistent with observations of the number of corrugations depending on density, as shown in the three snapshots for $\varphi_0=0.1$, $0.3$, $0.6$ (bottom), although the magnitude of $k$ seem to be overestimated.}
\label{Fig:app_shoot}
\end{figure}

To still allow for convergence, we apply the condition of null pressure~\eqref{eq:condelwP} slightly before the boundary, at $\lambda= \lambda_i + \varepsilon$, then let $\varepsilon \to 0$. The dispersion relation for the unstable mode converges toward its characteristic shape, as shown in Fig.~\ref{Fig:app_shoot}. However, the system becomes stiffer for larger $k$, and hence it was not possible to investigate the true extent of the apparent plateau at $s\approx 0.3$. On the other hand, the analysis gives a good qualitative understanding of how the instability depends on initial density, see also Fig.~\ref{Fig:app_shoot}. As no other parameter intervenes in the equations for the instability, and initial density does not affect the qualitative or scaling behavior to any significant extent, we argue that this analysis paints the instability discussed here as a robust property of dissipative blasts.

%\begin{figure}[h]
%\includegraphics*[height=120pt]{supp/kushnir_relatdisp.eps}\hspace*{3pt}
%\caption{Application of the shooting method to conservative blasts.}
%\label{Fig:app_kushnir}
%\end{figure}

\section*{Appendix B: Pressure anisotropy}

In the main text, we show that the usual assumption of conserved radial momentum~\cite{Oort1951} -- and the derived prediction $R(t)\sim t^{d/(d+1)}$ -- hinges on the assumption of vanishing momentum fluxes in the orthoradial direction (transfers between angular sectors). According to the classical argument, they are negligible because they are confined to the thin, dense peripheral shell of the blast. We propose that they may vanish exactly, even when the dense region has finite width, due to pressure in that region becoming purely radial. We discuss here in greater detail the consequences of pressure isotropy or anisotropy, both on the shape of the equations, and on predictions for the scaling of the blast radius $R(t)$.

\subsection*{Anisotropy and momentum fluxes}

If the pressure tensor $\bP(r)$ is isotropic, then we may write its divergence as a gradient
\[\nabla.\bP(r) = \nabla P(r) = \dr P(r) \er\]
If however the tensor is purely radial $\bP(r)=P(r) \er \otimes \er$ then we obtain the divergence of a vector field
\[\nabla.\bP(r) = \divg P(r) \er\]
To understand this, let us set $d=3$ and consider the small volume $d\mathcal V$ contained between two spherical caps located at $r$ and $r+dr$ and parametrized by angle $\theta \ll 1$. The surface area of the spherical cap located at $r$ is
\[2\pi (1-\cos \theta)r^2 = 4\pi \sin^2\left( \dfrac \theta 2 \right) r^2\] 
hence for small $\theta$
\[d\mathcal V = \pi \theta^2 r^2 dr \]
The force $\bF(r)\,d\mathcal V$ acting on this small volume is the integral of $\nabla.\bP$ over the volume, which is the pressure flux through the closed surface surrounding it. If the pressure is oriented in the radial direction, then it applies only on the spherical caps. Furthermore, it is here constant over each of them as it depends only on $r$, reducing the integral over one cap to the product of its constant pressure and its surface area :
\[\bF(r) d\mathcal V=\pi \theta^2 (P(r+dr) \,(r+dr)^2 - P(r) \,r^2) \er \]
and finally the force per unit volume which appears in the momentum equation is given in the limit $dr\to 0$ by
\[\bF(r) = \dfrac{1}{r^2} \dr \left(r^2 P(r) \right) \er = \left(\dr + \dfrac{2}r\right) P(r) \er\]
If however the pressure is isotropic, then there is also a contribution from the lateral surface of the volume, which is the same as for a cylinder with the corresponding mean radius, and height $dr \sin\theta$ :
\[2\pi \dfrac{(r \cos\theta + (r+dr)\cos\theta )}{2} dr \sin \theta \approx 2\pi\theta r dr \]
Furthermore, we may take $P(r)$ approximately constant over this lateral surface, and the resulting force must be projected on $\er$, which gives us an additional factor $ \sin(- \theta) \approx - \theta$ (the sign comes from the force pointing in the opposite direction). Hence, the force per unit volume is now given by
\[\bF(r).\er =  \dfrac{1}{r^2} \dr \left(r^2 P(r) \right) - \dfrac 2 r P(r) = \dr P(r) \]

\subsection*{Pseudo-MCS exponent}
As we have derived expressions for all fields in the cold region~\eqref{eq:coldprofiles} and for the latter's width~\eqref{eq:lami_approx}, we can check whether the assumption of purely radial pressure is well-grounded. At long times, we approximate $\lambda_c \approx 1$ i.e. we neglect the width of the cooling layer. Let us denote by  $\delta'$  the scaling exponent for $R(t)$ obtained in the case of an isotropic pressure within the cold region, and take Eq.~\eqref{eq:Blast.Hyd.Pcons} where we can now insert
\beqa \Pi(t) &=  \int_{0}^{\infty} n(r,t)\, u(r,t) \,r^{d-1}dr =  n_{RCP}\,\dot R\, R^{d}  \,(1-M_{RCP}^{-1})\,(1-\lambda_i)\eeqa
and
\beqa \int_{0}^{\infty} P(r,t)\, r^{d-2} dr &=  n_{i} \,\dot R^2 \, R^{d-1} \left( 1 -M_{RCP}^{-1}\right) \int_{\lambda_i}^{1} \left(  \lambda^{-d} (1-M_{RCP}) + M_{RCP}\right) d\lambda.  \eeqa
Using $n_{RCP}=n_0 M_{RCP}$ and simplifying $\Pi(t)/t$, we find
\beqa &\dt \Pi(\theta,t)=-(d-1) \int_{0}^{\infty} P(r,t)\, r^{d-2} dr\nonumber\\
&\delta'(d+1)-1 =- (d-1) \, \delta'-\delta' (1-M_{RCP}^{-1}) \dfrac{1-\lambda_i^{1-d}}{1-\lambda_i} \nonumber\eeqa
Hence, we prove that the scaling exponent differs from that imposed by radial momentum conservation $\delta=1/(d+1)$ (characterizing the MCS phase), as we find:
\beqa & \dfrac{1}{\delta'} = 1+ 2 d + \dfrac{M_{RCP}^{-1}}{\left(1-M_{RCP}^{-1}\right)^{1/d}-1 } \label{eq:valdelta}\\
&\lim_{n_0\to 0} \delta' = \dfrac{1}{d+1}, && \lim_{n_0 \to n_{RCP}} \delta' = \dfrac{1}{2d}\eeqa
We designate it as the ``pseudo-MCS exponent'' to reflect the fact that, while radial momentum is not exactly conserved and self-similarity is of the second kind (with a continuous dependence on microscopic parameters), it does recover the MCS exponent in the limit of low densities $\varphi_0 \to 0$.

\begin{figure}[h]
\includegraphics*[height=120pt]{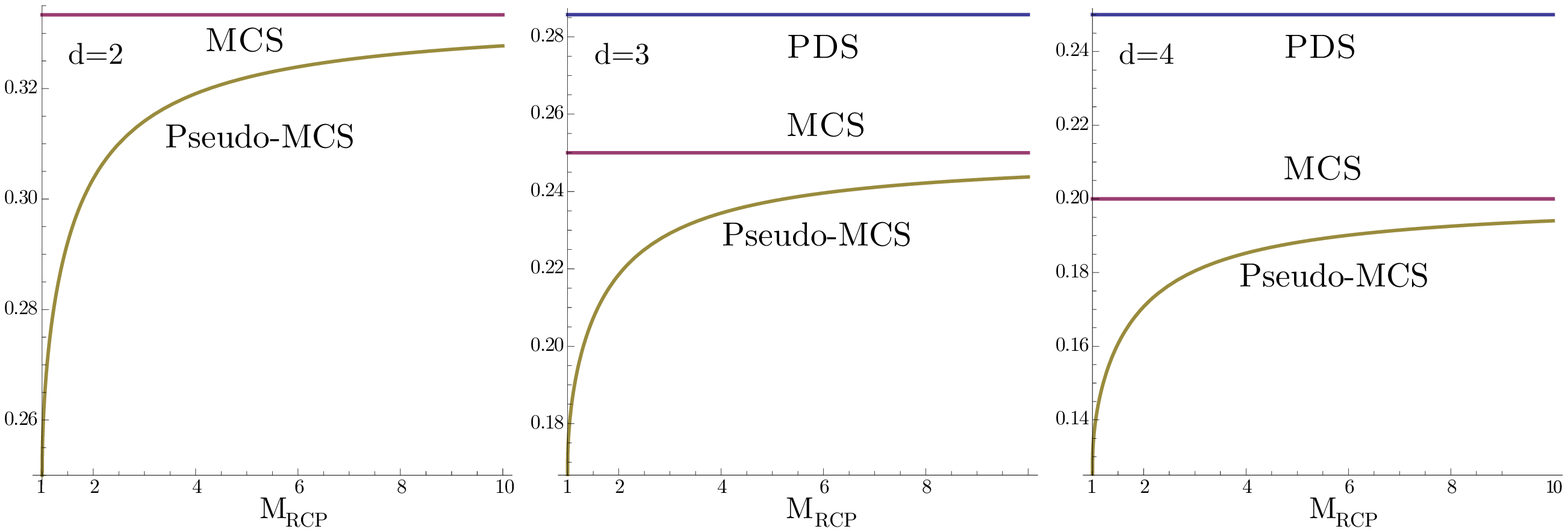}
\caption{Theoretical scaling exponent $\delta$ for the blast radius $R(t)\sim t^\delta$ in the Momentum-Conserving Snowplow (MCS) and Pressure-Driven Snowplow (PDS) regimes as defined in the main text, and ``pseudo-MCS'' exponent $\delta'$ defined in Eq.~\eqref{eq:valdelta}, as a function of the maximal compression $M_{RCP}=n_{RCP}/n_0$. Low initial densities $n_0$ -- as found in astrophysical systems -- entail high maximal compression, where the pseudo-MCS exponent tends to the MCS value. On the other hand, high initial densities can lead to a significant discrepancy, with a slower expansion of the blast. The dependence of the exponents on spatial dimension $d$ is illustrated here for $d=2$, 3 and 4; we recall that for $d=2$, the MCS and PDS exponents are equal (for hard spheres).}
\label{fig:anisoproof}
\end{figure}

We see on Fig.~\ref{fig:anisoproof} that the difference between the two exponents is around $10 \%$  for $\varphi_0=0.3$ in spatial dimension $d=3$, or $\varphi_0=0.5$ for $d=2$. It is clearly neglibible for initial densities considered in astrophyiscal systems, so that Oort's assumption is validated. However, this correction may become important in denser fluids, either granular or radatiave plasmas. Yet, measurements in our simulations in Fig.~\ref{fig:pconserv} have not allowed us to reject the hypothesis of asymptotic radial momentum conservation, which is exact only if pressure is truly anisotropic. As such anisotropy is well-attested in other granular systems, it remains a plausible ansatz for our solution.

\end{document}